\documentclass[twocolumn,nofootinbib,prd,aps,superscriptaddress,tightenlines]{revtex4}

\usepackage{slashed}
\usepackage{graphicx}
\usepackage{pstricks}
\usepackage{braket}
\usepackage{amsmath,amssymb}

\def\mreln{ \mathsf{M} }
\def\sreln{ R }

\def\N{N_c}
\def\eqref#1{{(\ref{#1})}}
\def\Tr{\textrm{Tr}}
\def\mc#1{{\mathcal{#1}}}
\def\D{\Delta}
\def\s{\sigma}
\def\a{\alpha}
\def\b{\beta}
\def\g{\gamma}
\def\l{\lambda}
\def\L{\Lambda}
\def\S{\Sigma}
\def\O{\Omega}
\def\ufont#1{\text{#1}}

{\count255=\time\divide\count255 by 60 \xdef\hourmin{\number\count255}
  \multiply\count255 by-60\advance\count255 by\time
  \xdef\hourmin{\hourmin:\ifnum\count255<10 0\fi\the\count255}}

\begin{document}

\title{A Lattice Test of $1/N_c$ Baryon Mass Relations}

\author{Elizabeth E.~Jenkins}
\affiliation{Department of Physics, University of California at San Diego,
  La Jolla, CA 92093}

\author{Aneesh V.~Manohar}
\affiliation{Department of Physics, University of California at San Diego,
  La Jolla, CA 92093}
  
\author{J.W. Negele}
\affiliation{Center for Theoretical Physics, Massachusetts Institute of Technology, Cambridge, MA 02139}

\author{Andr\'e Walker-Loud}
\affiliation{Department of Physics, College of William and Mary, Williamsburg, VA 23187-8795}

\begin{abstract}
$1/N_c$ baryon mass relations are compared with lattice simulations of baryon masses using different values of the light-quark masses, and hence different values of $SU(3)$ flavor-symmetry breaking.  The lattice data clearly display both the $1/N_c$ and $SU(3)$  flavor-symmetry breaking hierarchies. The validity of $1/N_c$ baryon mass relations derived without assuming approximate $SU(3)$ flavor-symmetry also can be tested by lattice data at very large values of the strange quark mass. The $1/N_c$ expansion constrains the form of discretization effects; these are suppressed by powers of $1/N_c$ by taking suitable combinations of masses.  This $1/N_c$ scaling is explicitly demonstrated in the present work.
\end{abstract}

\date{\today\quad\hourmin}

\maketitle

%
\section{Introduction}

The $1/N_c$ expansion of QCD~\cite{thooft} is a valuable tool for studying the nonperturbative dynamics of the strong interactions~\cite{witten,coleman}. In the limit $N_c \rightarrow \infty$, the baryon sector of QCD has an exact {\it contracted} $SU(2N_F)$ spin-flavor-symmetry~\cite{dm}.  For finite $N_c$, the contracted spin-flavor-symmetry is broken by effects suppressed by powers of $1/N_c$~\cite{dm,j}. The spin-flavor structure of the $1/N_c$ breaking terms is predicted at each order in the $1/N_c$ expansion~\cite{dm,j,djm1}.  The spin-flavor structure of many baryon properties have been derived in a systematic expansion in $1/N_c$~\cite{luty,carone,djm2,pirjol,schat,three}, and the results are in excellent agreement with experiment (for reviews, see~\cite{ejreview,amreview}).

One important application of the baryon $1/N_c$ expansion is to baryon masses~\cite{djm1,jl,h}.  By choosing appropriate linear combinations of the baryon masses, one can study coefficients of the baryon mass $1/N_c$ operator expansion with definite spin and flavor transformation properties. In the case of perturbative $SU(3)$ flavor-symmetry breaking, the $1/N_c$ analysis gives a hierarchy of baryon mass relations in powers of $1/N_c$ and the dimensionless $SU(3)$ breaking parameter $\epsilon \propto m_s/ \Lambda_\chi$~\cite{jl}.  The analysis in Ref.~\cite{jl} showed that the experimentally measured masses of the ground state octet and decuplet baryons exhibit the predicted hierarchy of the combined expansion in $1/N_c$ and $SU(3)$ flavor-symmetry breaking. The $1/N_c$ expansion also has been used to obtain very accurate predictions for the charm and bottom baryon masses~\cite{h} (to better than 10~MeV accuracy) which have been confirmed by recent experiments.  In addition, Ref.~\cite{djm1} derived baryon mass relations which only depend on the $1/N_c$ expansion and which are valid even if flavor $SU(3)$ is not a good approximate symmetry, i.e.\ for large values of the strange quark mass.

The predictions of the $1/N_c$ expansion for baryon masses are in excellent agreement with the experimental values. However, in comparing with the experimental data, one is restricted to only one value of $N_c$ and to the physical quark mass values.  Testing the predictions as a function of light-quark masses and $N_c$ is now possible with very accurate simulations of baryons using lattice QCD.

Tremendous progress in lattice QCD has been achieved recently in the simulation of baryon masses using different values of the light-quark masses.  Extrapolation of baryon masses on the lattice to the physical point has reproduced the experimental values at the $1-3\%$ level~\cite{Durr:2009ma}.  The lattice data, however, contain important additional information about the dependence of the baryon masses on the quark masses, which can be utilized.  Simulations of baryon masses have been performed as a function of $SU(3)$ flavor-symmetry breaking ranging from small perturbative flavor-symmetry breaking to large nonperturbative flavor-symmetry breaking.  There are also lattice simulations at different values of $N_c$ (though not for baryons) which are able to test  $N_c$ scaling rules~\cite{Teper:1998te,teper}.

In this paper, we show that existing lattice simulations (at $N_c=3$) of the ground state baryon masses already are sufficiently accurate to exhibit and test interesting features of the $1/N_c$ and $SU(3)$ flavor-symmetry breaking expansions.  Still more accurate simulations are needed to test the most suppressed mass combinations of the $1/N_c$ expansion, but  continued improvements in lattice simulations of baryon masses are expected in the short and long term, so it should eventually be possible to test these relations as well.   We discuss in this paper how present and future lattice data can be utilized to study the spin and flavor structure of baryon masses.  Although we do not focus on this point here, it should eventually be possible to test the $N_c$ scaling rules in the baryon sector by lattice simulations which vary the number of colors $N_c$.

An important observation is that the $1/N_c$ counting rules hold at finite lattice spacing, and so are respected by the lattice results \emph{including} the finite lattice spacing corrections dependent on the lattice spacing $a$. Thus the discretization corrections are constrained by the $1/N_c$ expansion.

Lattice computations of hadron masses are done at varying values of the light-quark masses $m_u$, $m_d$ and $m_s$, usually in the isospin limit $m_{u}=m_{d}\equiv m_{ud}$. The lattice results are then extrapolated to the physical values of $m_{ud}$.  It has not been possible to compute hadron masses for physical values of $m_{ud}$ yet due to the large computational time needed, since $\tau_{\text{comp}} \propto 1/m_{ud}^3$.%
\footnote{
One inverse power of the quark mass is from the conjugate-gradient inversion of the fermion Dirac operator, which scales 
with the condition number.  The additional inverse powers of the light-quark mass arise from estimations of (\textit{i}) the increased auto-correlation time of the HMC evolution as well as (\textit{ii}) the molecular dynamics step-size used to adjust the acceptance rate.   There are additional costs not counted here. For example, to keep the volume corrections exponentially small, one must work with $m_\pi L \gtrsim 4$, where $L$ is the spatial size of the lattice.} 
However, recently, with both algorithmic developments~\cite{Kennedy:2004ae} and large parallel computing machines, there are two groups simulating at or near the physical light-quark mass point~\cite{Durr:2009ma,Aoki:2008sm}.  

\begin{figure}
\includegraphics[width=5cm]{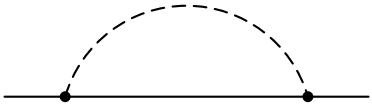}
\caption{One-loop correction to the baryon mass due to $\pi,K$ radiative corrections. \label{fig:loop}}
\end{figure}

The light-quark mass dependence of the hadron masses is determined by chiral perturbation theory.  There are nonanalytic in $m_q$ contributions from loop corrections which are calculable, as well as analytic terms which depend on low-energy constants (LECs) of the chiral Lagrangian.  In the baryon sector, the leading nonanalytic terms
are $m_q^{3/2}$ and $m_q^2 \ln m_q$.  The $m_q^{3/2}$ term, which arises from Fig.~\ref{fig:loop} and is proportional to $M^3_{\pi,K,\eta}/(16 \pi f^2)$ times axial couplings,
is found to be rather large, naively of order a few hundred MeV.  This term is absent for the pseudo-Goldstone boson masses, but is present for other mesons such as the vector mesons~\cite{wise}. 

A surprising feature of recent lattice results is that the baryon masses as a function of $M_{\pi,K,\eta}$ do not show a large $M^3$ dependence. Fits to the baryon masses also give much smaller values for the baryon axial couplings, and require that the $M^3\sim m_q^{3/2}$ term is almost completely cancelled by $m_q^2 \ln m_q$ and $m_q^2$ terms. This cancellation must be accidental at the currently accessible lattice quark masses, since the terms have different $m_q$ dependence.  An alternative conclusion is that lattice quark masses are too large for $SU(3)$ chiral perturbation theory to be valid, and that perturbative chiral behavior sets in only for much smaller quark masses than the strange quark mass.  This conclusion, however, fails to explain why $SU(3)$ flavor-symmetry is so evident in baryon phenomenology.

The $1/N_c$ expansion constrains the structure of baryon chiral perturbation theory~\cite{dm,djm1,djm2,flores}.  Chiral corrections have to respect the spin-flavor structure of the $1/N_c$ expansion, and this leads to some important restrictions on the form of the chiral loop corrections.  For example, the baryon mass, which is of order $N_c$~\cite{coleman,witten}, gets nonanalytic corrections proportional to $m_s^{3/2}$, which are large. This large nonanalytic contribution might lead one to expect that there should be large deviations from the Gell-Mann--Okubo mass relations, which were derived assuming that the mass operator was linear in $m_s$.  One can show, however, that the $N_c m_s^{3/2}$ terms are a spin-flavor singlet, and give the same contribution to octet and decuplet baryons, whereas the $m_s^{3/2}$ terms are a spin-singlet flavor-octet, and the $m_s^{3/2}/N_c$ terms are spin-singlet flavor-$\bf 27$.  Only the latter terms contribute to the Gell-Mann--Okubo mass combinations, so that deviations from these relations are a factor of $1/N_c^2$ smaller than naive expectation and consistent with experiment.  The small size of the Gell-Mann--Okubo relation was recently confirmed for a range of light-quark masses~\cite{Beane:2006pt}.  For QCD, the cancellation to order $1/N_c^2$ arises as a numerical cancellation between octet and decuplet intermediate states~\cite{Jenkins:1990jv}; to see the parametric form of the cancellation in $1/N_c$ requires computing the chiral corrections using the $SU(3)$ flavor representations of baryons containing $N_c$ quarks.

The mass relations of Ref.~\cite{jl} project the baryon masses onto different spin-flavor channels. By studying these mass relations as a function of $m_q$, one can investigate whether unexpected chiral behavior arises in a particular channel.

The organization of this paper is as follows:  In Sec.~II, the baryon mass relations of the $1/N_c$ expansion are summarized briefly, and in Sec.~III, the lattice simulation data are described.  Sec.~IV presents the results of a computation of $1/N_c$ mass combinations on the lattice for varying values of $SU(3)$ flavor-symmetry breaking. Sec.~V discusses the lattice analysis using heavy baryon chiral perturbation theory~\cite{Jenkins:1990jv}. Our conclusions are presented in Sec.~VI.

\section{$1/N_c$ Baryon Mass Relations}

The $1/N_c$ expansion of the baryon mass operator for perturbative $SU(3)$ flavor-symmetry breaking is\footnote{We use the notation and conventions of Ref.~\cite{jl}.}
\begin{eqnarray}\label{izero}
&&M = M^{1,0} + M^{ 8,0} +  M^{{27},0} + M^{{64},0}, \nonumber\\
&&M^{1,0} =  c_{(0)}^{1,0}\ \N \openone + c_{(2)}^{1,0}\ {1 \over \N }
J^2, \nonumber \\
&&M^{8,0} =  c_{(1)}^{8,0}\ T^8 + c_{(2)}^{8,0}\ {1 \over \N }
\{ J^i, G^{i8} \} + c_{(3)}^{8,0}\ {1 \over \N^2 } \{ J^2, T^8 \},
\nonumber \\
&&M^{{27},0} =   c_{(2)}^{27,0}\ {1 \over \N} \{ T^8, T^8 \}
+  c_{(3)}^{27,0}\ {1 \over \N^2 } \{ T^8, \{ J^i, G^{i8} \}\},
\nonumber \\
&&M^{{64},0} =   c_{(3)}^{64,0} {1 \over \N^2 } \{ T^8, \{ T^8, T^8
\}\},
\end{eqnarray}
where the superscript denotes the flavor $SU(3)$ representation and the spin $SU(2)$ representation of each term.
The $1/N_c$ expansion of the baryon mass operator contains eight independent operators, corresponding to the eight 
isospin multiplets of the ground state baryons.  
The arbitrary coefficients $c_{(n)}$ multiplying the $1/N_c$ operators are functions of $1/N_c$ and $SU(3)$ flavor-symmetry breaking.  Each coefficient is order $1$ at leading order in the $1/N_c$ expansion.    
The non-trivial content of the $1/N_c$ expansion is the $1/N_c$ suppression factors for the different terms in Eq.~(\ref{izero}).
Following Ref.~\cite{jl}, we define the parameter $\epsilon \propto m_s/\Lambda_\chi $, which is a dimensionless measure of flavor-$SU(3)$ symmetry breaking.  
$SU(3)$ flavor-symmetry breaking transforms as a flavor-octet, so the ${\bf 1}$, ${\bf 8}$, ${\bf 27}$ and ${\bf 64}$ operator expansions
are zeroth, first, second, and third order in $SU(3)$ flavor-symmetry breaking, respectively.  Thus, the coefficients $c^{8,0}$, $c^{27,0}$ and $c^{64,0}$ are naively order $\epsilon$, $\epsilon^2$ and $\epsilon^3$, respectively, in the $SU(3)$ flavor-symmetry breaking expansion.

The $1/N_c$ expansion for the baryon mass operator for $SU(2) \times U(1)$ flavor-symmetry~\cite{djm1} is given by
\begin{eqnarray}\label{su2u1}
&&M = \N \openone + K + {1 \over \N} I^2 + {1 \over \N} J^2 + {1 \over \N} K^2 \nonumber\\
&&+ {1 \over \N^2} I^2 K + {1 \over \N^2} J^2 K
+ {1 \over \N^2} K^3,
\end{eqnarray}
where each operator is accompanied by an unknown coefficient which has been suppressed for simplicity.  Each coefficient is order unity at leading order in the $1/N_c$ expansion.  The operator $K \equiv N_s/2$, where $N_s$ is the $s$-quark number operator.  Again, there are eight independent operators in the $1/N_c$ expansion corresponding to the masses of the eight isomultiplets of the ground state baryons.

\begin{table*}
\caption{Mass combinations $\mreln_1$--$\mreln_8$ from Ref.~\cite{jl} and $\mreln_A$--$\mreln_D$ from Ref.~\cite{djm1}.  The coefficients and orders in $1/\N$ and perturbative $SU(3)$ flavor
symmetry breaking $\epsilon$ are given for mass combinations $\mreln_1$--$\mreln_8$.  Combinations $\mreln_A$--$\mreln_D$ are obtained at order $1/N_c^2$ assuming only isospin flavor-symmetry. \label{table:relns}}
\begin{eqnarray*}
\renewcommand\arraystretch{1.5}
\setlength\arraycolsep{0.2cm}
\begin{array}{|c|c|c|c|c|c|}
\hline \hline
\text{Label} & \text{Operator} & \text{Coefficient} & \text{Mass\  Combination}  & 1/\N & SU(3) \\
\hline
 \mreln_1 & \openone &160\, \N \ c^{1,0}_{(0)} & 25(2N +\Lambda+3\Sigma+2\Xi) 
 	-4(4\Delta +3\Sigma^* +2\Xi^* +\Omega) & \N
& 1 \\
\mreln_2 & J^2 &-120\, \frac{1}{\N} \ c^{1,0}_{(2)} & 5(2N +\Lambda+3\Sigma+2\Xi) 
	-4(4\Delta +3\Sigma^* +2\Xi^* +\Omega)& 1/\N & 1 \\
\mreln_3 & T^8 & 20 \sqrt{3}\, \epsilon \ c^{8,0}_{(1)} & 5(6N +\Lambda-3\Sigma -4\Xi) 
	-2(2\Delta -\Xi^* -\Omega)& 1 & \epsilon \\
\mreln_4 & \{ J^i, G^{i8} \} & -5 \sqrt{3}\, \frac{1}{\N}\ \epsilon \ c^{8,0}_{(2)} & N +\Lambda -3\Sigma +\Xi & 1/\N & \epsilon \\
\mreln_5 & \{ J^2, T^8 \} &30 \sqrt{3}\, \frac{1}{\N^2}\ \epsilon \ c^{8,0}_{(3)} &(-2N +3\Lambda-9\Sigma + 8\Xi) 
	+2(2\Delta -\Xi^* -\Omega) & 1/\N^2 & \epsilon \\
\mreln_6 & \{ T^8, T^8 \} & 126\, \frac{1}{\N} \ \epsilon^2 \ c^{27,0}_{(2)} & 35(2N -3\Lambda-\Sigma+2\Xi) 
	-4(4\Delta -5\Sigma^* -2\Xi^* +3\Omega) & 1/\N & 
 \epsilon^2 \\
\mreln_7 & \{T^8, J^i G^{i8} \} & -63\,\frac{1}{\N^2} \ \epsilon^2 \ c^{27,0}_{(3)} & 7(2N -3\Lambda-\Sigma+ 2\Xi) 
	-2(4\Delta -5\Sigma^* -2\Xi^* +3\Omega) &
1/\N^2 & \epsilon^2 \\
\mreln_8 & \{ T^8, \{ T^8, T^8 \} \} & 9 \sqrt{3}\,\frac{1}{\N^2} \ \epsilon^3 \ c^{64,0}_{(3)} & \Delta - 3 \Sigma^* + 3 \Xi^* - \Omega & 1/\N^2 & \epsilon^3 \\[5pt]
\hline
\mreln_A &  &  & \left( \Sigma^*  - \Sigma \right)- \left( \Xi^*  - \Xi \right)   & 
1/\N^2 & - \\
\mreln_B &  &  &  {1 \over 3} \left( \Sigma + 2 \Sigma^* \right) - \Lambda - {2 \over 3} \left( \Delta - N \right)  & 1/\N^2 & - \\
\mreln_C &  &  & -{1 \over 4} \left( 2 N - 3  \Lambda -  \Sigma +2 \Xi \right)  +{1 \over 4} \left( \Delta -\Sigma^* - \Xi^* + \Omega \right)  & 1/\N^2 & - \\
\mreln_D &  &  &  -{1 \over 2} \left( \Delta - 3  \Sigma^* +3  \Xi^*-  \Omega \right)  & 1/\N^2 & - \\
\hline\hline
\end{array}
\end{eqnarray*}
\end{table*}

The $1/N_c$ expansion continues to hold at finite lattice spacing, so that expansions of the form Eqs.~(\ref{izero})
and (\ref{su2u1}) are valid with coefficients that depend on the lattice spacing $a$.\footnote{Rotational symmetry is broken down to a discrete cubic symmetry, so that irreducible representations of the spin $SU(2)$ group are replaced by irreducible representations of the cubic group.} Thus, the order $N_c$ discretization correction is universal, and has the same value for all the octet and decuplet baryons. This term can be eliminated by studying baryon mass ratios or differences. The largest term that produces different discretization effects for the octet baryons is the $T^8$ term in $M^{8,0}$, or equivalently the $K$ term in Eq.~(\ref{su2u1}), which results in an $\mathcal{O}(1)$ mass splitting that is suppressed by one factor of $1/N_c$ relative to the leading mass term $N_c\openone$.

The twelve $1/N_c$ mass relations that we study are tabulated in Table~\ref{table:relns}. The eight mass relations $\mreln_1$--$\mreln_8$ correspond to the first eight mass combinations in Table~II of Ref.~\cite{jl}.  These mass relations are the mass combinations which occur at definite orders in $1/N_c$ and perturbative $SU(3)$ flavor-symmetry breaking; each relation picks out a particular coefficient in Eq.~(\ref{izero}).  The mass combinations in Table~\ref{table:relns} correspond to the coefficients listed in the table.  There are two mass combinations, $\mreln_1$ and $\mreln_2$, which are $SU(3)$ singlets and which occur at orders $N_c$ and $1/N_c$, respectively.
There are three flavor-octet mass combinations, $\mreln_3$, $\mreln_4$ and $\mreln_5$, which are proportional
to one factor of $SU(3)$ symmetry breaking $\epsilon$ and which occur at orders $1$, $1/N_c$ and $1/N_c^2$, respectively.  There are two flavor-$\bf 27$ mass combinations, $\mreln_6$ and $\mreln_7$, which occur at second order in $SU(3)$ breaking $\epsilon^2$ and at orders $1/N_c$ and $1/N_c^2$ in the $1/N_c$ expansion.  Finally, there is one mass combination $\mreln_8$ which is suppressed by three powers of $SU(3)$ breaking $\epsilon^3$ and by $1/N_c^2$ in the $1/N_c$ expansion.  The additional four mass combinations $\mreln_A$--$\mreln_D$ correspond to the first four mass relations of Ref.~\cite{djm1}.  These baryon mass combinations are each order $1/N_c^2$ in the $1/N_c$ expansion.  Only three of the four combinations $\mreln_A$--$\mreln_D$ are linearly independent, corresponding to the fact that there are three independent operators at order $1/N_c^2$ in Eq.~(\ref{su2u1}).
These mass relations were derived assuming only isospin flavor-symmetry, and so do not assume approximate $SU(3)$ flavor-symmetry and are valid even for very large $SU(3)$ flavor-symmetry breaking.  Note that the $1/N_c^2$ mass relations $\mreln_A$--$\mreln_D$ are related to the $1/N_c^2$ relations $\mreln_5$, $\mreln_7$ and $\mreln_8$ by
\begin{eqnarray}\label{mam5}
\mreln_A &=& {1 \over {10}} \mreln_5 + {1 \over {70}} \mreln_7 - {2 \over 7} \mreln_8, \nonumber \\
\mreln_B &=& - { 1 \over {15}} \mreln_5 + {4 \over {105}} \mreln_7 - {2 \over {21}} \mreln_8, \nonumber \\
\mreln_C &=& -{1 \over {28}} \left( \mreln_7 + \mreln_8 \right), \nonumber \\
\mreln_D &=& -{1 \over 2} \mreln_8 .
\end{eqnarray}   

Each mass relation defines a mass combination of the octet and decuplet masses $M_i$ of the form
\begin{equation}
\mreln = \sum_i c_i M_i \ .
\end{equation}
In this work, isospin breaking is neglected, so each $M_i$ denotes the average mass of a baryon isomultiplet. The normalization of each mass relation is arbitrary, and depends on a particular choice of normalization for coefficients in the Hamiltonian, i.e.\ $c_i \to \lambda c_i$ defines another mass relation with the same spin and flavor quantum numbers. To remove the rescaling ambiguity, Ref.~\cite{jl} used the accuracy defined by
\begin{eqnarray}\label{eq:Areln}
A \equiv {{\sum_i c_i M_i } \over {{\frac 1 2 \sum_i  | c_i | M_i }}} .
\end{eqnarray}
to quantify the fractional error of a given mass relation. Here we use a related quantity, the scale invariant mass combination
\begin{eqnarray}\label{eq:Rreln}
\sreln \equiv {{\sum_i c_i M_i } \over {{\sum_i  | c_i | }}} .
\end{eqnarray}
Dividing by $\sum_i  | c_i |$ instead of by ${1 \over 2} \sum_i  | c_i | M_i$ avoids mixing different flavor representations via the denominator factor. The rescaled relations $R_{1}$--$R_8$ and $R_{A}$--$R_{D}$ have dimensions of mass.

In our numerical analysis, we shall use the dimensionless variable
\begin{eqnarray}
\epsilon &=& \frac{M_K^2-M_\pi^2}{\Lambda_\chi^2}
\label{edef}
\end{eqnarray}
as a measure of $SU(3)$ breaking, where $\Lambda_\chi \sim 4 \pi f = 1$~GeV~\cite{georgi} is the scale of chiral symmetry breaking.

\section{Lattice Simulation}
In this work, we use the results of the recent LHPC spectrum calculation~\cite{WalkerLoud:2008bp} to explore the mass combinations of the $1/N_c$ expansion.  The LHP Collaboration utilized a mixed-action lattice calculation with domain-wall~\cite{Kaplan:1992bt,Shamir:1993zy,Furman:1994ky} valence propagators computed with the Asqtad improved~\cite{Orginos:1998ue,Orginos:1999cr} dynamical MILC gauge ensembles~\cite{Bernard:2001av,Bazavov:2009bb}.%
\footnote{The strange quark and many of the light-quark propagators were computed by the NPLQCD Collaboration~\cite{Beane:2008dv}.} 
The calculation was performed at one lattice spacing with $a\sim 0.125$~\ufont{fm}, and a fixed spatial volume $L\sim 2.5$~\ufont{fm}.  The pion and kaon masses used in Ref.~\cite{WalkerLoud:2008bp} are $\{M_\pi,M_K\} = \{293,586\}$, $\{356,604\}$, $\{496,647\}$, $\{597,686\}$, $\{689,729\}$ and $\{758,758\}$~\ufont{MeV}, respectively, on the \texttt{m007}, \texttt{m010}, \texttt{m020}, \texttt{m030}, \texttt{m040}, and \texttt{m050} ensembles, where the labels denote the light-quark masses in lattice units.%
\footnote{The mass $M_\eta$ is defined at this order by the Gell-Mann--Okubo formula $M_\eta^2 = \frac 4 3 M_K^2 -\frac13 M_\pi^2$.}
In the dynamical ensembles and the computation of the valence propagators, the strange quark was held fixed near its physical value.  (In fact the strange quark was $\sim25\%$ too large~\cite{Aubin:2004ck}.)  For further details of the calculation, we refer the reader to Ref.~\cite{WalkerLoud:2008bp}.  

Using the \textit{bootstrap resampled} lattice data, we determine the 12 mass combinations of Table~\ref{table:relns} on each ensemble.  The results are collected in Table~\ref{table:lattRelns}.  These  results were determined with the absolute scale of $a^{-1}=1588$~\ufont{MeV} on all coarse ensembles, where the scale used in Ref.~\cite{WalkerLoud:2008bp}
was determined from heavy-quark spectroscopy.  We have additionally determined the mass combinations using the smoothed 
values of $r_1/a$, where $r_1$ is determined on each different ensemble from the heavy-quark potential with 
$r_1^2 F(r_1) = 1$~\cite{Aubin:2004wf}.  The values of $a^{-1}$ determined in this 
way range from $\{1597, 1590, 1614, 1621, 1628, 1634\}$~\ufont{MeV}  from the lightest to heaviest quark mass. 
These two scale-setting methods are in 
good agreement, as shown in the next section. 
  
\begin{table*}
\caption{\label{table:lattRelns} Values of mass combinations $\mreln_i$ in \ufont{GeV} using the scale setting $a^{-1}=1.588$~\ufont{GeV}.}
\begin{eqnarray*}
\begin{array}{|c|rrrrrr|}
\hline\hline
M_K^2 - M_\pi^2 [\ufont{GeV}^2]&0.2579 & 0.2380 & 0.1718 & 0.1141 & 0.0574& 0\\
\hline
\mreln_1 [\ufont{GeV}]& 192(1)& 197(1)& 211(1)& 221(2)& 237(2)& 242(1) \\
\mreln_2 [\ufont{GeV}]& -12.2(5) & -13.3(6) & -10.9(4) & -10.6(6) & -7.8(6)& -8.2(4) \\
\mreln_3 [\ufont{GeV}]& -8.07(20) & -7.15(15) & -4.40(08) & -2.81(09) & -1.28(04)& 0\\
\mreln_4 [\ufont{GeV}]&  -0.214(10)& -0.181(7)& -0.099(5)& -0.056(6)& -0.022(3)& 0\\
\mreln_5 [\ufont{GeV}]& 0.29(7)& 0.35(8)& 0.13(3)& 0.14(5)& 0.04(2)& 0 \\
\mreln_6 [\ufont{GeV}]& -1.05(36)& -0.25(26)& -0.003(71)& 0.15(14)& -0.02(1)& 0 \\
\mreln_7 [\ufont{GeV}]& -0.30(12)& -0.05(11)& 0.005(24)& 0.08(7)& -0.003(4)& 0 \\
\mreln_8 [\ufont{GeV}]& 0.02(1)& 0.012(09)& -0.001(1)& 0.015(12)& 0.00004(6)& 0 \\
\hline
\mreln_A [\ufont{GeV}]& 0.0004(58)& 0.018(7)& 0.001(3)& 0.004(4)& -0.0003(19)& 0 \\
\mreln_B [\ufont{GeV}]& -0.022(10)& -0.018(10)& -0.0004(30)& -0.003(3)& 0.0000(13)& 0 \\
\mreln_C [\ufont{GeV}]& 0.010(4)& 0.002(4)& -0.0002(8)& -0.003(3)& 0.0001(1)& 0 \\
\mreln_D [\ufont{GeV}]& -0.012(5)& -0.0061(45)& 0.0004(6)& -0.0076(58)& -0.00002(3)& 0 \\
\hline\hline
\end{array}
\end{eqnarray*}
\end{table*}

%
\section{Comparison with Lattice Data \label{sec:comparison}}

\begin{figure}
\includegraphics[bb=40 400 490 720,width=8cm]{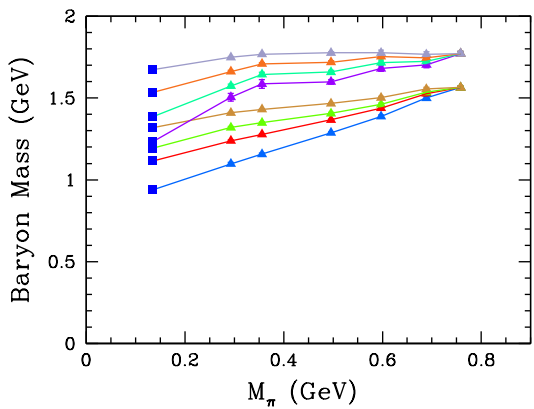}
\caption{Octet and decuplet baryon lattice masses as a function of $M_\pi$. The blue squares are the PDG values and the triangles are LHP Collaboration lattice data simulated with fixed $m_s$ and varying $m_{ud}$.  The two right-most points simulate the flavor-singlet baryon masses in the exact $SU(3)$ flavor-symmetry limit $M_\pi = M_K$.  Note that the value of $M_\pi$ at the $SU(3)$ flavor-symmetry point is quite large, $M_\pi =758$~MeV. The (PDG) baryons from bottom to top are $N,\Lambda,\Sigma,\Delta,\Xi,\Sigma^*,\Xi^*$ and $\Omega$. \label{fig:mbt}}
\end{figure}

A plot of the lattice baryon masses as a function of $M_\pi$ is given in Fig.~\ref{fig:mbt}.
The triangles are the lattice data, and the blue squares are the Particle Data Group (PDG) values.\footnote{Note that $M_K$ varies slightly for the different $M_\pi$ values and is not exactly equal to the physical $M_K$.} 
The usual way of studying the data is to fit to 
the individual baryon masses. The $1/N_c$ analysis shows that it is useful to study the mass combinations in Table~\ref{table:relns} instead, as these combinations have definite orders in $1/N_c$ and $SU(3)$ symmetry breaking.

In this section, we present a series of plots of the 12 scale invariant baryon mass combinations $R_1$--$R_8$ and $R_A$--$R_D$ as a function of $SU(3)$ flavor-symmetry breaking.  In each plot, the red triangles and green circles are the lattice data, and the errors on the mass combinations have been computed using the \textit{bootstrap} data sets of the octet and decuplet masses, taking advantage of the full correlations in  the lattice data. The red triangles and green circles  are the lattice results using the two different scale-setting methods.  The red points are determined with the absolute scale setting $a^{-1}=1588$~\ufont{MeV}, and the green points are determined with the smoothed $r_1/a$ values~\cite{Aubin:2004wf}.  The blue square is obtained using the physical baryon and meson masses from the PDG~\cite{pdg}, with the error bar computed using experimental uncertainties on the isospin averaged masses. For most of the plots, the error on the blue point is smaller than size of the point, and is not visible.

The horizontal axis is $M_K^2-M_\pi^2$ in units of $\text{(GeV)}^2$, which is a measure of $SU(3)$ flavor-symmetry breaking, and the vertical axis is the mass combination in MeV. For mass combinations proportional to powers of $SU(3)$ breaking, we plot both the mass combination and the mass combination divided by the appropriate power of $\epsilon$, using the definition Eq.~(\ref{edef}) for $\epsilon$, so that the units remain MeV.

The average $O(N_c)$ mass of the octet and decuplet baryons is $\sim\!1000$~MeV.  Naive power counting with $N_c=3$ implies order $N_c$ masses  are $\sim\! 1000$~MeV, order $1$ masses are $\sim\!300$~MeV,
order $1/N_c$ masses are $\sim\!100$~MeV,  and order $1/N_c^2$ masses are $\sim\!30$~MeV.  This $1/N_c$ hierarchy is evident in the PDG and lattice data for the baryon mass combinations $R_1$--$R_8$ and $R_A$--$R_D$ below.

\begin{figure}
\includegraphics[bb=40 400 490 720,width=8cm]{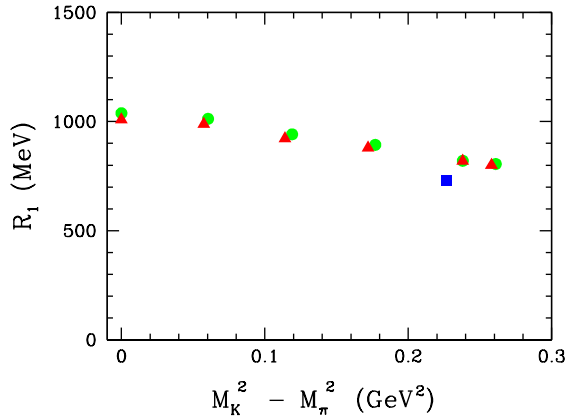}
\caption{$R_1$ as a function of $(M_K^2 - M_\pi^2)$. The red triangles and green circles are lattice results with two different scale setting methods (see text), and the blue square is using the PDG values for the hadron masses. $R_1$ is the average ${\cal O}(N_c)$ mass of the ground state baryons. \label{fig:r1}}
\end{figure}

Figure~\ref{fig:r1} plots the first relation $R_1$, which is the scale invariant mass corresponding to the $N_c \openone$ operator in the $1/N_c$ expansion.  This mass relation is order $1000$~MeV over the range of $SU(3)$ flavor-symmetry breaking simulated by the LHP Collaboration and is seen to be fairly independent of $SU(3)$ breaking as expected.  The deviation of the physical PDG point from the lattice data is presumably due to the larger than physical strange quark mass
of the simulation.  In principle, this relation also is susceptible to finite discretization errors, beginning at $\mc{O}(a^2)$, although these are expected to be small for this data set~\cite{WalkerLoud:2008bp}.

\begin{figure}
\includegraphics[bb=40 400 490 720,width=8cm]{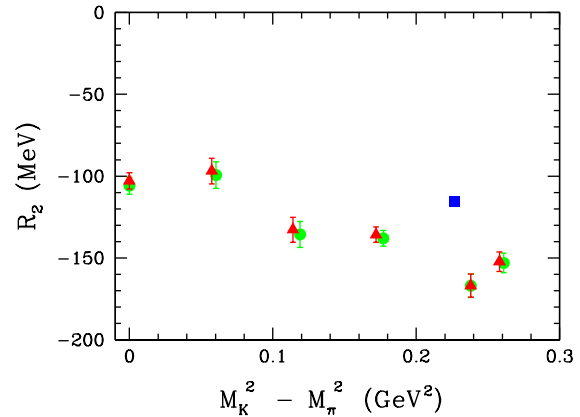}
\caption{$R_2$ as a function of $(M_K^2 - M_\pi^2)$. The red triangles and green circles are lattice results with two different scale setting methods (see text), and the blue square is using the PDG values for the hadron masses.  $R_2$ is minus the ${\cal O}(1/N_c)$ hyperfine mass splitting between the spin-$3/2$ decuplet and the spin-$1/2$ octet baryons. \label{fig:r2}}
\end{figure}

Figure~\ref{fig:r2} plots the second relation $R_2$, which corresponds to the $J^2/N_c$ operator in the $1/N_c$ expansion.  The mass combination $R_2$ is clearly suppressed relative to the average ground state baryon mass $R_1$. This hyperfine splitting is predicted to be order $1/\N$, and so of order $100$~MeV in magnitude. The lattice data allow us to study this mass relation as a function of $SU(3)$ breaking. Notice that the relation works independently of $SU(3)$ breaking as predicted theoretically, and that $R_2$ does not vanish at the $SU(3)$ symmetry point $M_K^2=M_\pi^2$. The numerical results clearly show that the suppression of $R_2$ is a consequence of $1/N_c$, and not due to a hidden flavor breaking suppression factor.

\begin{figure}
\includegraphics[bb=40 150 490 720,width=8cm]{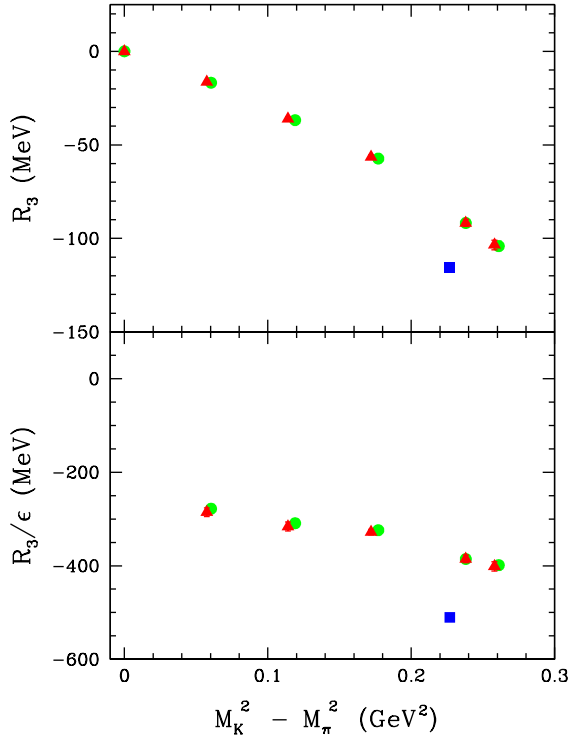}
\caption{$R_3$ and $R_3/\epsilon$ as a function of $(M_K^2 - M_\pi^2)$. $R_3$ is ${\cal O}(\epsilon)$ at leading order
in the $1/N_c$ and $SU(3)$ flavor-symmetry breaking expansions, and $R_3/\epsilon$ is ${\cal O}(1)$. \label{fig:r3}}
\end{figure}

Figure~\ref{fig:r3} plots the third relation $R_3$, which is of order $\epsilon$. The relation vanishes as $M_K^2-M_\pi^2 \to 0$, as can be seen in the upper panel. The lower panel divides the mass relation by $\epsilon$, and shows that $R_3/\epsilon$ is an $O(1)$ mass in the $1/N_c$ expansion, or $\sim 300$~MeV in magnitude.  Notice that $R_3/\epsilon$ does not vanish in the $SU(3)$ symmetry limit.  The $SU(3)$ symmetry point of the lattice data is omitted in the $R_3/\epsilon$ plot, as it involves dividing by zero.

\begin{figure}
\includegraphics[bb=40 150 490 720,width=8cm]{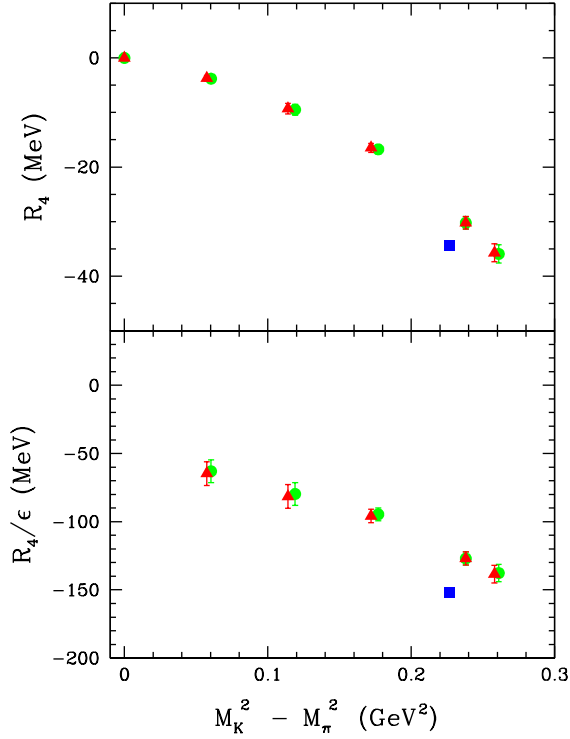}
\caption{$R_4$ and $R_4/\epsilon$ as a function of $(M_K^2 - M_\pi^2)$. $R_4$ is ${\cal O}(\epsilon/N_c)$ at leading order, and $R_4/\epsilon$ is ${\cal O}(1/N_c)$.\label{fig:r4}}
\end{figure}

Figure~\ref{fig:r4} plots $R_4$ and $R_4/\epsilon$, which are of order $\epsilon/\N$ and $1/\N$, respectively. The mass combinations $R_3$ and $R_4$ are both order $\epsilon$, but $R_4$ is suppressed relative to $R_3$ by an additional factor of $1/N_c$. This  suppression is clearly visible in the numerical values which are $\sim 100$~MeV in magnitude.  $R_4$ is seen to vanish in the $SU(3)$ symmetry limit, whereas $R_4/\epsilon$ does not vanish at the $SU(3)$ symmetric point. Both $R_3/\epsilon$ and $R_4/\epsilon$ display some dependence on $SU(3)$ symmetry breaking, which implies that there is $\epsilon$ dependence in the coefficients $c_{(1)}^{8,0}$ and $c_{(2)}^{8,0}$, respectively.

\begin{figure}
\includegraphics[bb=40 150 490 720,width=8cm]{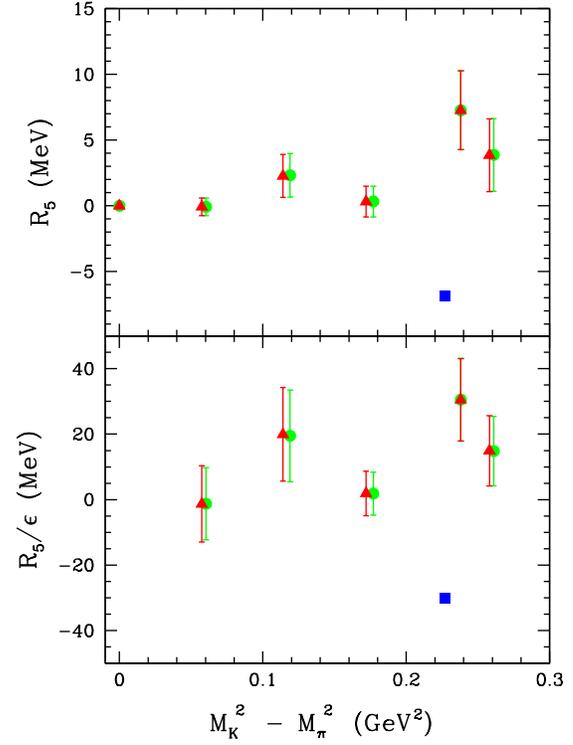}
\caption{$R_5$ and $R_5/\epsilon$ as a function of $(M_K^2 - M_\pi^2)$.  $R_5$ is ${\cal O}(\epsilon/N_c^2)$ at leading order, and $R_5/\epsilon$ is ${\cal O}(1/N_c^2)$. \label{fig:r5}}
\end{figure}

Figure~\ref{fig:r5} plots $R_5$ and $R_5/\epsilon$, which are of order $\epsilon/\N^2$ and $1/\N^2$, respectively. The mass relation $R_5/\epsilon$ is consistent with being a $1/N_c^2$ mass of magnitude $\sim 30$~MeV.   The lattice data points now have relatively large errors, however.

\begin{figure}
\includegraphics[bb=40 150 490 720,width=8cm]{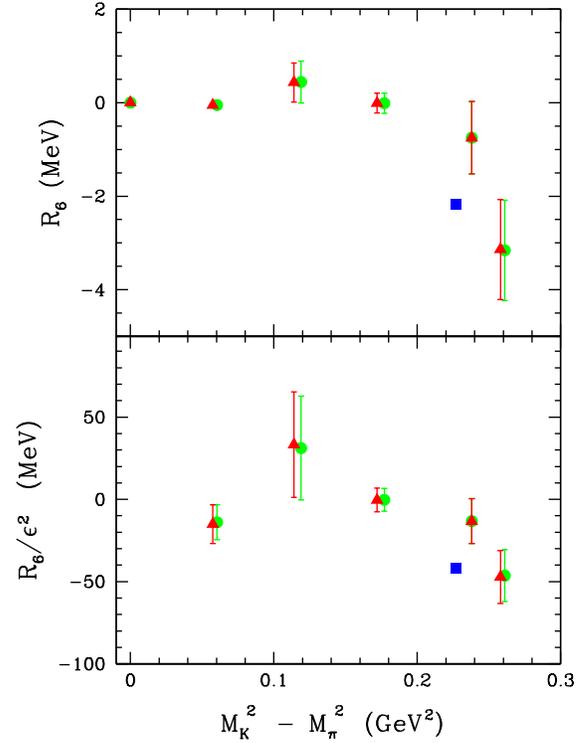}
\caption{$R_6$ and $R_6/\epsilon^2$ as a function of $(M_K^2 - M_\pi^2)$.  $R_6$ is ${\cal O}(\epsilon^2/N_c)$, and $R_6/\epsilon^2$ is ${\cal O}(1/N_c)$. \label{fig:r6}}
\end{figure}

\begin{figure}
\includegraphics[bb=40 150 490 720,width=8cm]{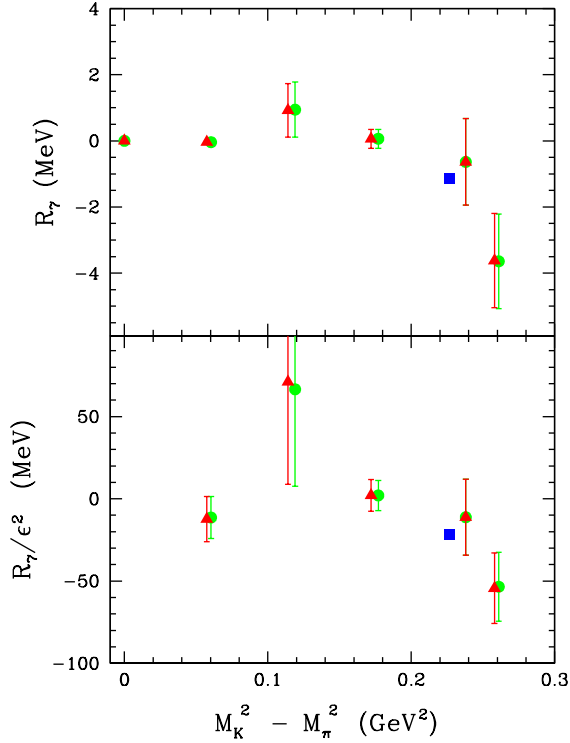}
\caption{$R_7$ and $R_7/\epsilon^2$ as a function of $(M_K^2 - M_\pi^2)$. $R_7$ is ${\cal O}(\epsilon^2/N_c^2)$, and $R_7/\epsilon^2$ is ${\cal O}(1/N_c^2)$. \label{fig:r7}}
\end{figure}

\begin{figure}
\includegraphics[bb=40 150 490 720,width=8cm]{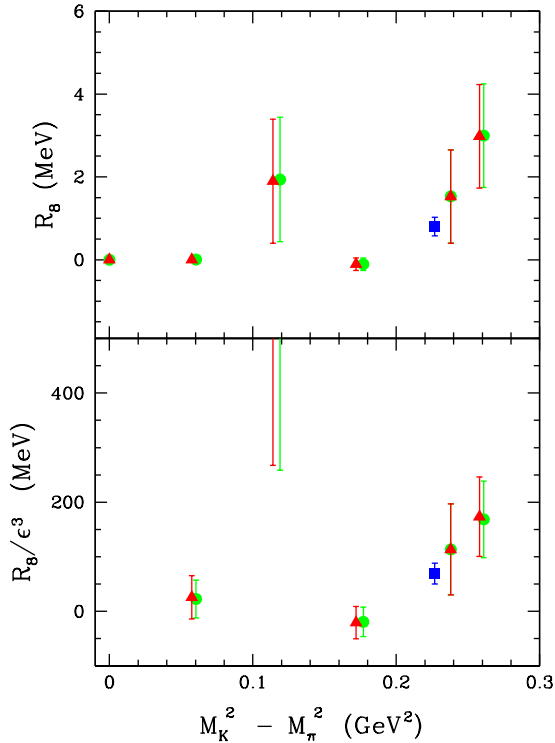}
\caption{$R_8$ and $R_8/\epsilon^3$ as a function of $(M_K^2 - M_\pi^2)$. $R_8$ is ${\cal O}(\epsilon^3/N_c^2)$, and $R_8/\epsilon^3$ is ${\cal O}(1/N_c^2)$. \label{fig:r8}}
\end{figure}

$R_{6}$, $R_7$ and $R_8$ are plotted in Figs.~\ref{fig:r6}, ~\ref{fig:r7} and~\ref{fig:r8}, respectively, for completeness. In each case, the physical point (blue square) has the expected size, and the lattice results are compatible with $1/N_c$ and $\epsilon$ power counting expectations, but the error bars are now rather large.  It would be very interesting to have more precise lattice data for the $R_6$ and $R_7$ mass relations because the
leading $SU(3)$ flavor-symmetry breaking contribution to the flavor-${\bf 27}$ mass splittings in chiral perturbation theory is ${\cal O}(\epsilon^{3/2})$ rather than the naive ${\cal O}(\epsilon^2)$ of second order perturbative flavor-symmetry breaking.  With very precise data, as in Ref.~\cite{Beane:2009ky}, it would be
possible to test this prediction of chiral perturbation theory.  

\begin{figure}
\includegraphics[bb=40 400 490 720,width=8cm]{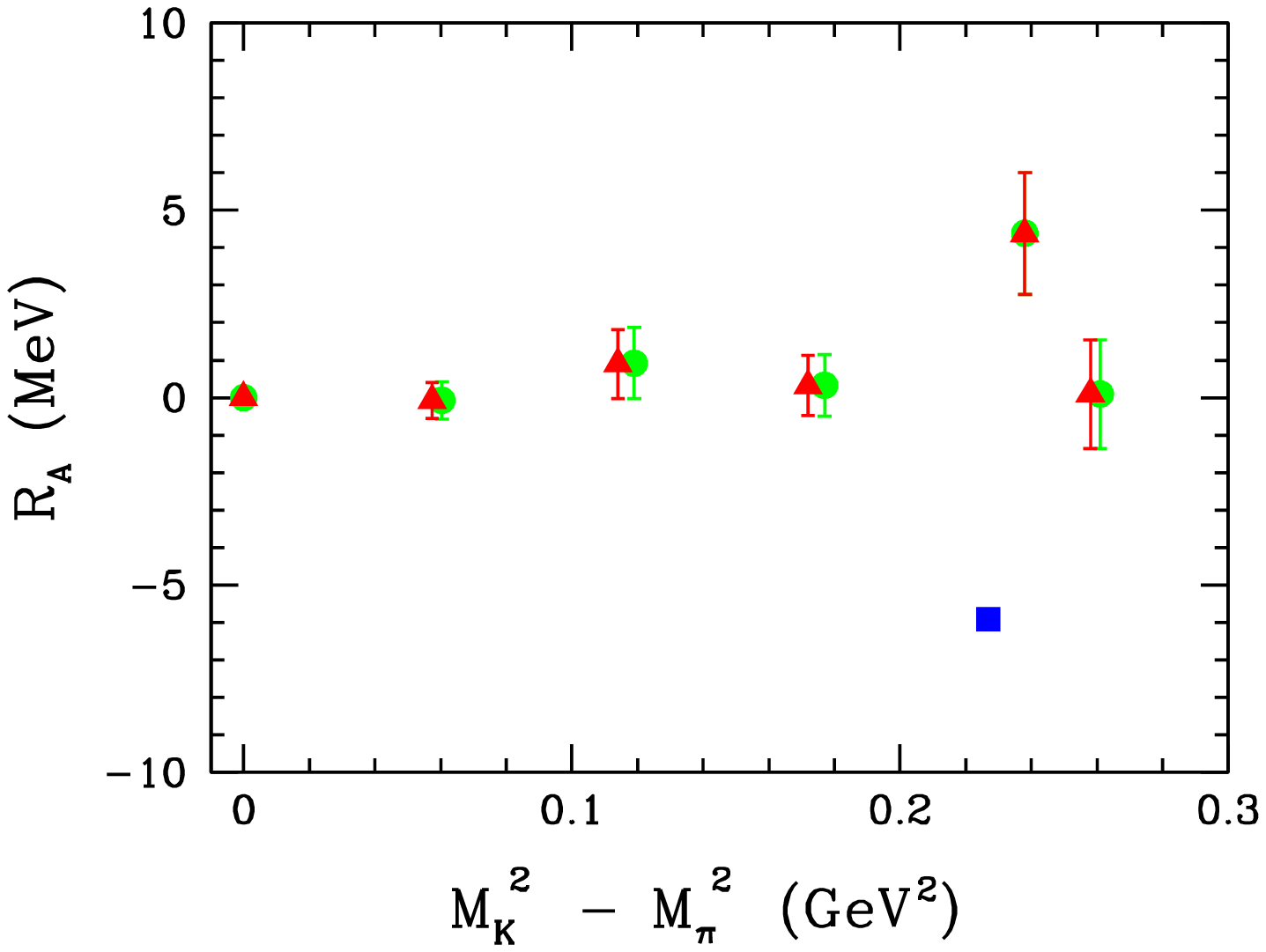}
\caption{$R_A$ as a function of $(M_K^2 - M_\pi^2)$.  $R_A$ is ${\cal O}(1/N_c^2)$. \label{fig:djm1}}
\end{figure}

\begin{figure}
\includegraphics[bb=40 400 490 720,width=8cm]{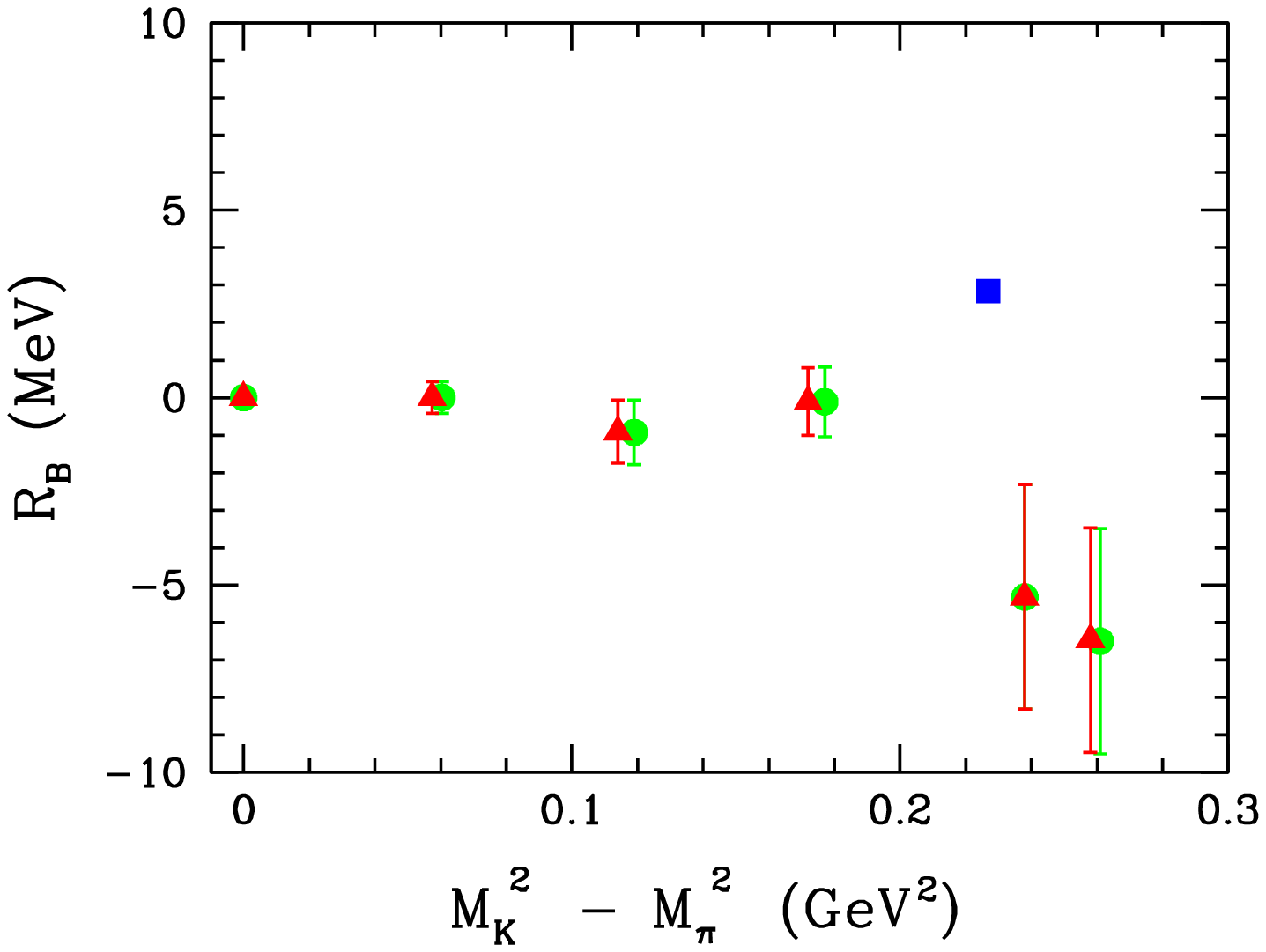}
\caption{$R_B$ as a function of $(M_K^2 - M_\pi^2)$.   $R_B$ is ${\cal O}(1/N_c^2)$. \label{fig:djm2}}
\end{figure}

\begin{figure}
\includegraphics[bb=40 400 490 720,width=8cm]{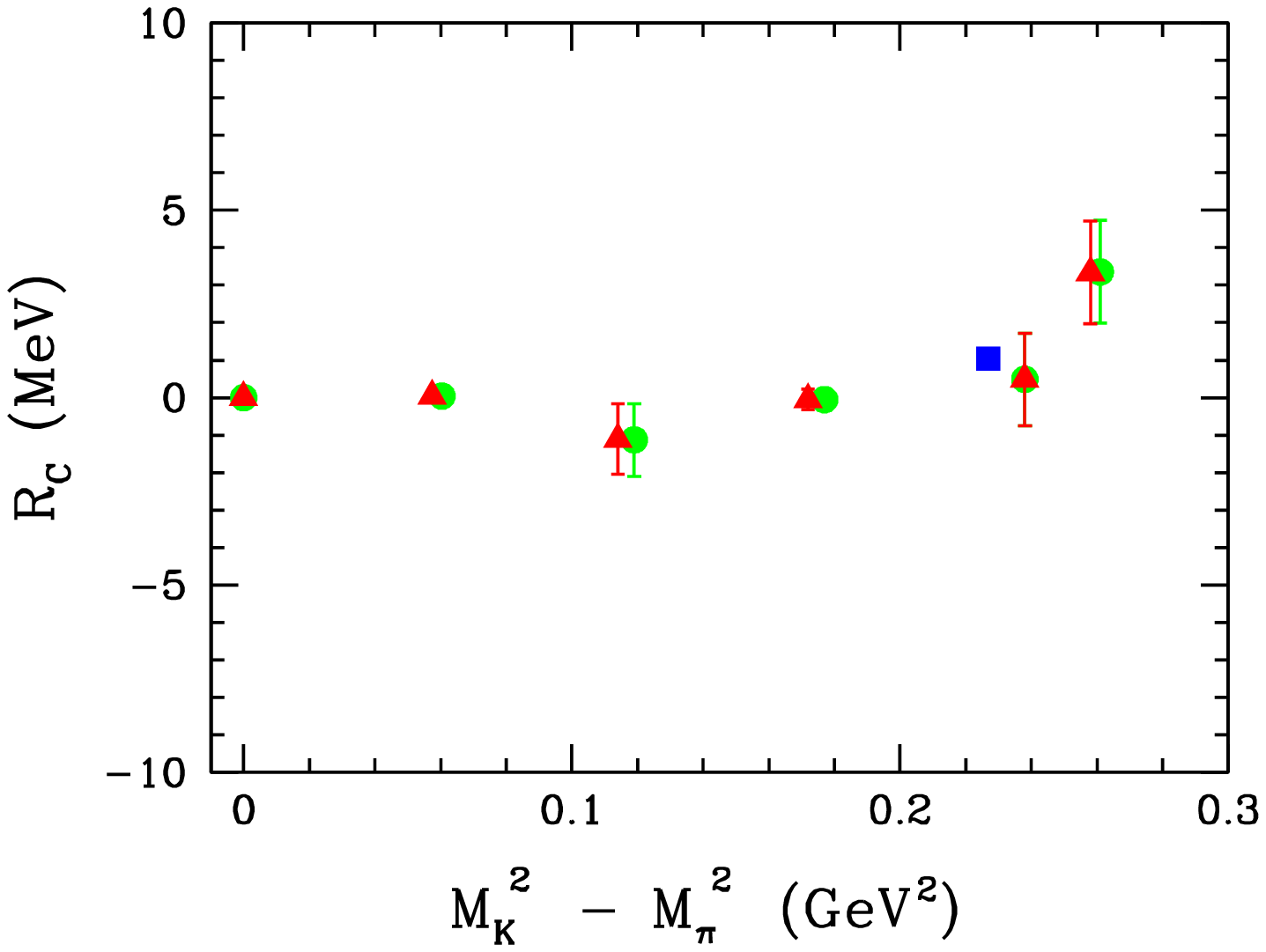}
\caption{$R_C$ as a function of $(M_K^2 - M_\pi^2)$.   $R_C$ is ${\cal O}(1/N_c^2)$. \label{fig:djm3}}
\end{figure}

\begin{figure}
\includegraphics[bb=40 400 490 720,width=8cm]{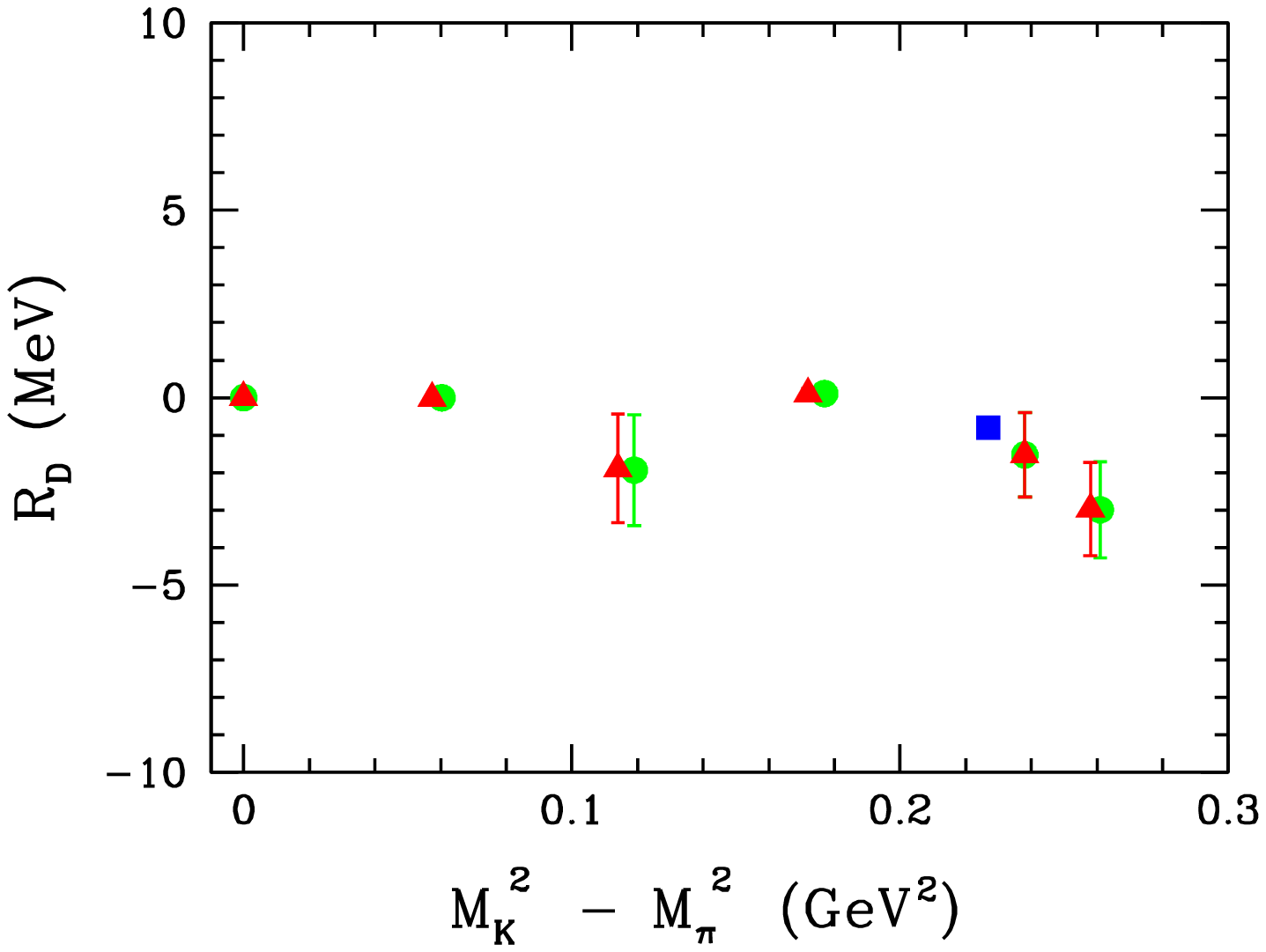}
\caption{$R_D$ as a function of $(M_K^2 - M_\pi^2)$.   $R_D$ is ${\cal O}(1/N_c^2)$. \label{fig:djm4}}
\end{figure}

The last four plots, Figs.~\ref{fig:djm1}--\ref{fig:djm4},
are the relations $R_A$--$R_D$, which are of order $1/N_c^2$ and assume only $SU(2) \times U(1)$ flavor-symmetry. From the plots, it is evident that $R_A$--$R_D$ are satisfied irrespective of the value of $SU(3)$ breaking.  $R_A$--$R_D$ are all predicted to be ${\cal O}(1/N_c^2)$ masses, or approximately $30$~MeV.  The figures show that $R_A$--$R_D$ are significantly smaller than $30$~MeV.  This suppression can be understood from the perturbative $SU(3)$ flavor-symmetry breaking analysis, since the lattice data varies $\epsilon$ only over the range
$[0,0.2579]$ and does not contain large nonperturbative values of $\epsilon$.  From Eq.~(\ref{mam5}), one sees that for small values
of $\epsilon$, $\mreln_A$ and $\mreln_B$ are ${\cal O}(\epsilon/N_c^2)$ since they are linear combinations containing $\mreln_5$;
$\mreln_C$ is ${\cal O}(\epsilon^2/N_c^2)$ since it is a linear combination containing $\mreln_7$; and $\mreln_D$ is ${\cal O}(\epsilon^3/N_c^2)$ since it is proportional to $\mreln_8$.    
It would be interesting to see how well relations $R_A$--$R_D$ work at even \emph{larger} values of $SU(3)$ breaking extending to very large values for which the $SU(3)$ flavor-symmetry is completely broken. In the limit $m_s \to \infty$, relation $R_A$ goes over to the mass relation for heavy-quark baryons, $\Sigma^*_Q-\Sigma_Q=\Xi^*_Q-\Xi_Q^\prime$, so it would be interesting to look at the crossover 
from light-quark mass behavior proportional to $m_s$ to heavy-quark behavior in which the hyperfine splitting is proportional to $1/m_Q$.  Similarly, in the limit $m_s \to \infty$, relation $R_B$ goes over to the heavy-quark baryon mass relation $\left[ {1 \over 3} \left( \Sigma_Q + 2 \Sigma_Q^* \right) - \Lambda_Q \right] - {2 \over 3} \left( \Delta - N \right)$, which is independent of the heavy-quark mass.  The crossover from $\sim m_s$ behavior to mass-independence in the $m_s \to \infty$ limit also would be interesting to observe.  In addition, it
would be interesting to determine the value of $\epsilon$ at which the perturbative $SU(3)$ flavor-symmetry breaking analysis fails.  Finally, an extended range for $\epsilon \propto M_K^2 - M_\pi^2$ would allow one to see whether relations $R_A$--$R_D$ remain independent of $\epsilon$ even for large values of $\epsilon$.

\section{Heavy Baryon $\chi$PT Analysis}
As mentioned in the introduction, a recent surprise from current lattice light-quark spectrum calculations is that the baryon masses do not show a large $M_{\pi,K,\eta}^3$ behavior.  
In fact, the baryon masses display an unexpected (nearly) linear in $M_\pi$ scaling for a large range of pion masses~\cite{WalkerLoud:2008bp}.  In the case of the nucleon, this finding was verified in all current $2+1$ flavor lattice calculations of the nucleon mass~\cite{WalkerLoud:2008pj}.  These findings pose an interesting theoretical puzzle and indicate the presence of large cancellations to the baryon spectrum from the different orders in the chiral expansion.  Further, recent $SU(3)$ chiral extrapolations of the baryon masses~\cite{WalkerLoud:2008pj,Ishikawa:2009vc}, carried out using the next-to-leading order (NLO) extrapolation formulae, have found that the axial couplings, $D$, $F$, $C$ and $H$, when left as free parameters, are returned from the analysis with values significantly different from those expected based upon phenomenology~\cite{FloresMendieta:1998ii} or the recent lattice determination~\cite{Lin:2007ap}.
It is not clear at this point whether these findings are due to a breakdown of $SU(3)$ heavy baryon $\chi$PT in this mass range, or, for example, the necessity of including the next-to-next-to-leading order (NNLO) terms in the analysis.

In this section, we perform an $SU(3)$ heavy baryon $\chi$PT extrapolation analysis of the various mass relations $\mreln_i$ presented in Table~\ref{table:lattRelns}.  This allows us both to extrapolate our results to the physical $\{M_\pi,M_K\}$ limit and to begin exploring the combined $1/N_c$ and $SU(3)$ expansion utilizing precision lattice data.  An important question to explore is whether this combined expansion displays significantly improved convergence properties to the $SU(3)$ extrapolations performed previously.  Additionally, because the different mass relations are sensitive to the different coefficients of the baryon mass expansion in Eq.~\eqref{izero}, we can directly extrapolate different orders in the $SU(3)$ chiral expansion.  For example, as mentioned above, the leading contribution to $\mreln_6$ and $\mreln_7$ scales as $\mathcal{O}(\epsilon^{3/2})$ in heavy baryon $\chi$PT.  Finding a definitive signal of this scaling would be a significant confirmation of the nonanalytic light-quark mass behavior predicted by chiral perturbation theory.

The light and strange quark mass dependence is described by $SU(3)$ heavy baryon chiral perturbation theory (HB$\chi$PT)~\cite{Jenkins:1990jv}.  The baryon chiral Lagrangian is given by
\begin{align}\label{eq:HBLag}
\mathcal{L} =&\ \Tr \left( \bar{B} i v \cdot {\cal D} B \right)
	- \bar{T}^\mu \left[ i v \cdot {\cal D} -\D_0 \right] T_\mu
\nonumber\\&
	+b_D \Tr \left( \bar{B} \{\mc{M}_+,  B\} \right)
	+b_F \Tr \left( \bar{B} [\mc{M}_+,  B]  \right)
\nonumber\\&
	+b_0 \Tr \left( \bar{B} B \right) \Tr \left( \mc{M}_+ \right)
\nonumber\\&
	+\gamma_M\, \bar{T}^\mu \mc{M}_+ T_\mu
	-\bar{\s}\, \bar{T}^\mu T_\mu \Tr \left( \mc{M}_+ \right)
\nonumber\\&
	+2D\, \Tr \left( \bar{B} S^\mu \{\mc{A}_\mu, B\} \right)
	+2F\, \Tr \left( \bar{B} S^\mu [\mc{A}_\mu, B] \right)
\nonumber\\&
	+2\mc{H}\, \bar{T}^\mu S_\nu \mc{A}^\nu T_\mu
	+\mc{C}\, \left( \bar{T}^\mu \mc{A}_\mu B + \bar{B} \mc{A}_\mu T^\mu \right)\, ,
\end{align}
where $B$ and $T_\mu$ are the octet and decuplet fields, respectively; $v_\mu$ is the four-velocity of the baryon and $S_\mu$ is the spin-vector.  The decuplet--octet mass splitting in the chiral limit is $\D_0$, and $\mc{M}_+$ is the mass spurion defined by
\begin{align}
	&\mc{M}_+ = 
	\xi^\dagger m_Q \xi^\dagger + \xi m_Q^\dagger \xi &
\end{align}
in terms of the light-quark mass matrix
\begin{equation}
m_Q = \begin{pmatrix}
	m_u &&\\
	&m_d&\\
	&&m_s
	\end{pmatrix}\, ,
\end{equation}
and $\xi = e^{i \Pi/ f_\pi}$.
The covariant derivative is
\begin{equation}
	{\cal D}_\mu B = \partial_\mu B + [\mc{V}_\mu , B]\, ,
\end{equation}
and the vector and axial fields are
\begin{align}
	&\mc{V}_\mu = \frac{1}{2} \left( \xi \partial_\mu \xi^\dagger + \xi^\dagger \partial_\mu \xi \right)\, ,&\\
	&\mc{A}_\mu = \frac{i}{2} \left( \xi \partial_\mu \xi^\dagger - \xi^\dagger \partial_\mu \xi \right)\, .&
\end{align}
The masses of the octet and decuplet baryons were first determined in Ref.~\cite{Jenkins:1991ts} to NLO, $\mc{O}(M_{K,\pi}^3)$.  The octet baryon masses were later determined to NNLO, $\mc{O}(M_{K,\pi}^4)$ in Refs.~\cite{Lebed:1994gt,Borasoy:1996bx,WalkerLoud:2004hf}, and the decuplet baryons to NNLO in Refs.~\cite{Lebed:1993yu,Tiburzi:2004rh}.  

The lattice calculation of baryon masses used for this work was a mixed-action calculation~\cite{WalkerLoud:2008bp}.  Therefore, the calculation and the corresponding low-energy effective field theory for the baryons are inherently partially quenched.  The partially quenched Lagrangian for heavy baryons was determined by Chen and Savage~\cite{Chen:2001yi} through NLO.  The extension of the partially quenched Lagrangian to the mixed-action Lagrangian, which includes corrections from the lattice spacing, also is known~\cite{Tiburzi:2005is,Chen:2007ug}, through NNLO~\cite{Chen:2007ug,WalkerLoud:2004hf,Tiburzi:2004rh}.  At NLO, the corrections from the mixed-action are straightforward, amounting to corrections to the mixed valence-sea meson masses~\cite{Orginos:2007tw} appearing in the one-loop self-energy corrections, Fig.~\ref{fig:loop}, as well as the partially quenched hairpin corrections~\cite{Chen:2001yi}.  A rigorous extrapolation to the continuum limit requires multiple lattice spacings, but the lattice data used in this work was performed at a single lattice spacing (albeit with the size of the discretization corrections expected to be small, see Ref.~\cite{WalkerLoud:2008bp} for details).  However, one can make a qualitative estimate of the size of the discretization errors by comparing the extrapolations performed with the continuum extrapolation formulae and the mixed-action formulae. We now perform these chiral extrapolations.

\subsection{Variable projected $\chi^2$ minimization}
To construct the $\chi^2$ for the light-quark mass extrapolation, we use the \textit{bootstrap resampled} lattice data to form the covariance matrix,
\begin{align}
&C_{q,q^\prime} = \frac{1}{N_{bs}} \sum_{bs}^{N_{bs}} \Big( \mreln_q[bs] - \bar{\mreln}_q \Big) 
	\Big( \mreln_{q^\prime}[bs] - \bar{\mreln}_{q^\prime} \Big)\, ,&
\end{align}
where $q$ is a supermass index which runs over both the mass relations $\mreln_1$--$\mreln_7$ as well as the different lattice ensembles \texttt{m007}--\texttt{m050}.%
\footnote{To the order we work in $SU(3)$ heavy baryon $\chi$PT, the relation $\mreln_8$ is exactly zero.  In the partially quenched theory, this mass relation exactly vanishes at NLO, and at NNLO, there are residual partially quenched effects which do not exactly cancel~\cite{Tiburzi:2004rh}.} 
We then construct the $\chi^2$,
\begin{align}
&\chi^2 = \sum_{q,q^\prime} \Big( \bar{\mreln}_q - f(\mreln_q,\lambda) \Big) C_{q,q^\prime}^{-1} \Big( \bar{\mreln}_{q^\prime} - f(\mreln_{q^\prime},\lambda) \Big)\, ,&
\end{align}
where $f(\mreln_q,\lambda)$ is the HB$\chi$PT extrapolation formula for the mass relation $\mreln_q$ and depends upon the low-energy constants $\vec{\l} \equiv \{M_0,\D_0, b_D,b_F, b_0, \bar{\s},\g_M,D,F,\mc{C},\mc{H},\dots\}$. At NLO, the baryon mass extrapolation formulae are linearly dependent upon all the $\l_i$, except for $D$, $F$, $\mc{C}$ and $\mc{H}$, but they  are linearly dependent upon $\mc{C}^2$ and $\mc{H}^2$.  Therefore, to perform the numerical minimization, we first perform a linear least squares fit on all the linear LECs, $\vec{\l}_{lin} \equiv \{M_0,\D_0, b_D,b_F, b_0, \bar{\s},\g_M,\mc{C}^2,\mc{H}^2\}$, solving for the low-energy constants as functions of $D$ and $F$, and then perform the numerical minimization.  This procedure is known as the variable projection (VarPro) method~\cite{VarPro}.

We first perform the $\chi^2$ minimization using relations $\mreln_1$--$\mreln_7$ and the NLO extrapolation formulae.  In Tables~\ref{table:NLOR1R7} and \ref{table:NLOMAR1R7}, we collect the results of both the continuum extrapolation and the mixed-action extrapolation.  In these tables, the partially quenched/mixed-action low-energy constants $\a_M$, $\b_M$, $\s_M$, $\g_M$ and $\bar{\s}_M$ are related to the LECs of $SU(3)$ HB$\chi$PT of Eq.~\eqref{eq:HBLag} by
\begin{align}
&b_D = \frac{1}{4}\left( \a_M - 2\b_M \right)\, ,& 
	&\g_M = \g_M^{(PQ)}\, ,&
\\
&b_F = \frac{1}{2} \left( 5\a_M + 2\b_M \right)\, ,&
	&\bar{\s}_M = \bar{\s}_M^{(PQ)}\, ,&
\\
&b_0 = \s_M + \frac{1}{6}\a_M + \frac{2}{3}\b_M\, .&
\end{align}
Then, for example, $\tilde{\a}_M = \a_M / B_0$, where at leading order $M_K^2 = B_0 (m_s + m_{ud})$.  In both cases, we perform the fit for four different ranges of the quark masses.  The first fit includes only the lightest two mass points, while each successive fit includes an additional mass point.  The results of the mixed-action and continuum fits are fairly consistent, as can be seen in Tables~\ref{table:NLOR1R7} and \ref{table:NLOMAR1R7}.
\begin{table*}
\caption{\label{table:NLOR1R7} Fit to Mass Relations $R_1$--$R_7$ using NLO continuum HB$\chi$PT with a variation projection (VarPro) method.}
\begin{eqnarray*}
\begin{array}{|c|cc|ccccccccc|}
\hline\hline
\textrm{Fit Range}& D& F& M_0& \Delta_0& \tilde{\a}_M& \tilde{\b}_M& \tilde{\g}_M& \tilde{\s}& \tilde{\bar{\s}}& C^2& H^2\\
&&& [\ufont{GeV}]& [\ufont{GeV}]& [\ufont{GeV}] ^{-1}& [\ufont{GeV}] ^{-1}& [\ufont{GeV}]^{-1}& [\ufont{GeV}]^{-1} & [\ufont{GeV}] ^{-1}&& \\
\hline
\texttt{m007}-\texttt{m010}& 
	0.04(28)& 0.14(6)& 0.999(49)& 0.21(16)& -0.58(7)& -0.42(16)& 0.68(37)& -0.10(6)& -0.36(31)& 0.01(12)& 0.32(55) \\
\texttt{m007}-\texttt{m020}& 
	0.17(8)& 0.07(5)& 0.972(18)& 0.48(5)& -0.67(2)& -0.81(5)& 0.39(12)& -0.11(2)& 0.09(9)& 0.21(3)& 0.09(19)\\
\texttt{m007}-\texttt{m030}& 
	0.21(5)& 0.08(3)& 0.972(13)& 0.44(5)& -0.67(2)& -0.76(3)& 0.49(12)& -0.12(1)& -0.00(8)& 0.15(2)& 0.18(18)\\
\texttt{m007}-\texttt{m040}& 
	0.20(5)& 0.09(2)& 0.960(11)& 0.43(04)& -0.67(1)& -0.73(3)& 0.62(8)& -0.13(1)& -0.05(6)& 0.14(2)& 0.35(13) \\
\hline\hline
\end{array}
\end{eqnarray*}
\end{table*}
\begin{table*}
\caption{\label{table:NLOMAR1R7} Fit to Mass Relations $R_1$--$R_7$ using NLO mixed-action HB$\chi$PT with a variation projection (VarPro) method.}
\begin{eqnarray*}
\begin{array}{|c|cc|ccccccccc|}
\hline\hline
\textrm{Fit Range}& D& F& M_0& \Delta_0& \tilde{\a}_M& \tilde{\b}_M& \tilde{\g}_M& \tilde{\s}& \tilde{\bar{\s}}& C^2& H^2\\
&&& [\ufont{GeV}]& [\ufont{GeV}]& [\ufont{GeV}] ^{-1}& [\ufont{GeV}] ^{-1}& [\ufont{GeV}]^{-1}& [\ufont{GeV}]^{-1} & [\ufont{GeV}] ^{-1}&& \\
\hline
\texttt{m007}-\texttt{m010}& 
	 0.31(9)& 0.26(5)& 0.941(42)& 0.242(73)& -1.29(1)& -0.46(1)& 3.8(9)& -0.35(5)& -1.8(4)& -0.043(9)& 4.3(1.1)\\
\texttt{m007}-\texttt{m020}& 
	0.36(3)& 0.19(2)& 0.989(15)& 0.365(30)& -1.03(1)& -0.70(1)& 0.66(22)& -0.24(2)& -0.22(11)& -0.009(4)& 0.21(26)\\
\texttt{m007}-\texttt{m030}& 
	0.35(2)& 0.16(2)& 1.001(11)& 0.368(28)& -0.92(1)& -0.71(1)& 0.58(21)& -0.20(1)& -0.15(10)& -0.007(4)& 0.11(25) \\
\texttt{m007}-\texttt{m040}& 
	 0.34(2)& 0.15(1)& 0.995(8)& 0.376(26)& -0.87(1)& -0.70(1)& 0.69(18)& -0.20(1)& -0.16(8)& -0.001(2)& 0.21(21)\\
\hline\hline
\end{array}
\end{eqnarray*}
\end{table*}

\begin{figure}[htb]
\includegraphics[width=8cm]{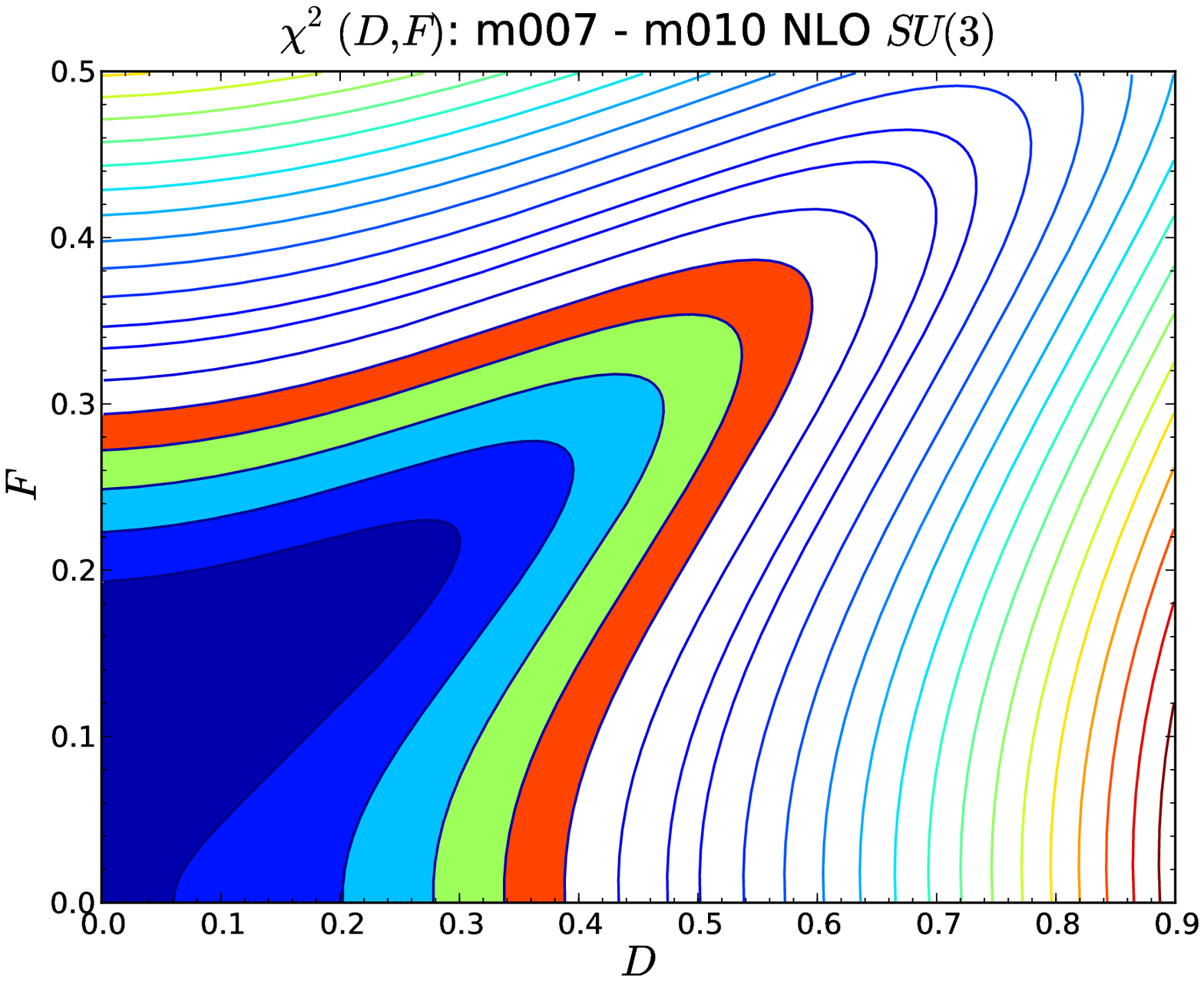}
\caption{Contour plot of $\chi^2(D,F)$ constructed using a continuum $SU(3)$ HB$\chi$PT $\chi^2$ fit to the \texttt{m007}--\texttt{m010} data sets.  The dark (blue) inner region represents $\chi^2 \lesssim \chi^2_{min} + 2.30$, the $\pm1\sigma$ confidence region for two fit parameters, $D$ and $F$.  Each successive $n^{\rm th}$ contour represents the $\pm n \sigma$ confidence region.
\label{fig:chisq007010}}
\end{figure}

\begin{figure}[htb]
\includegraphics[width=8cm]{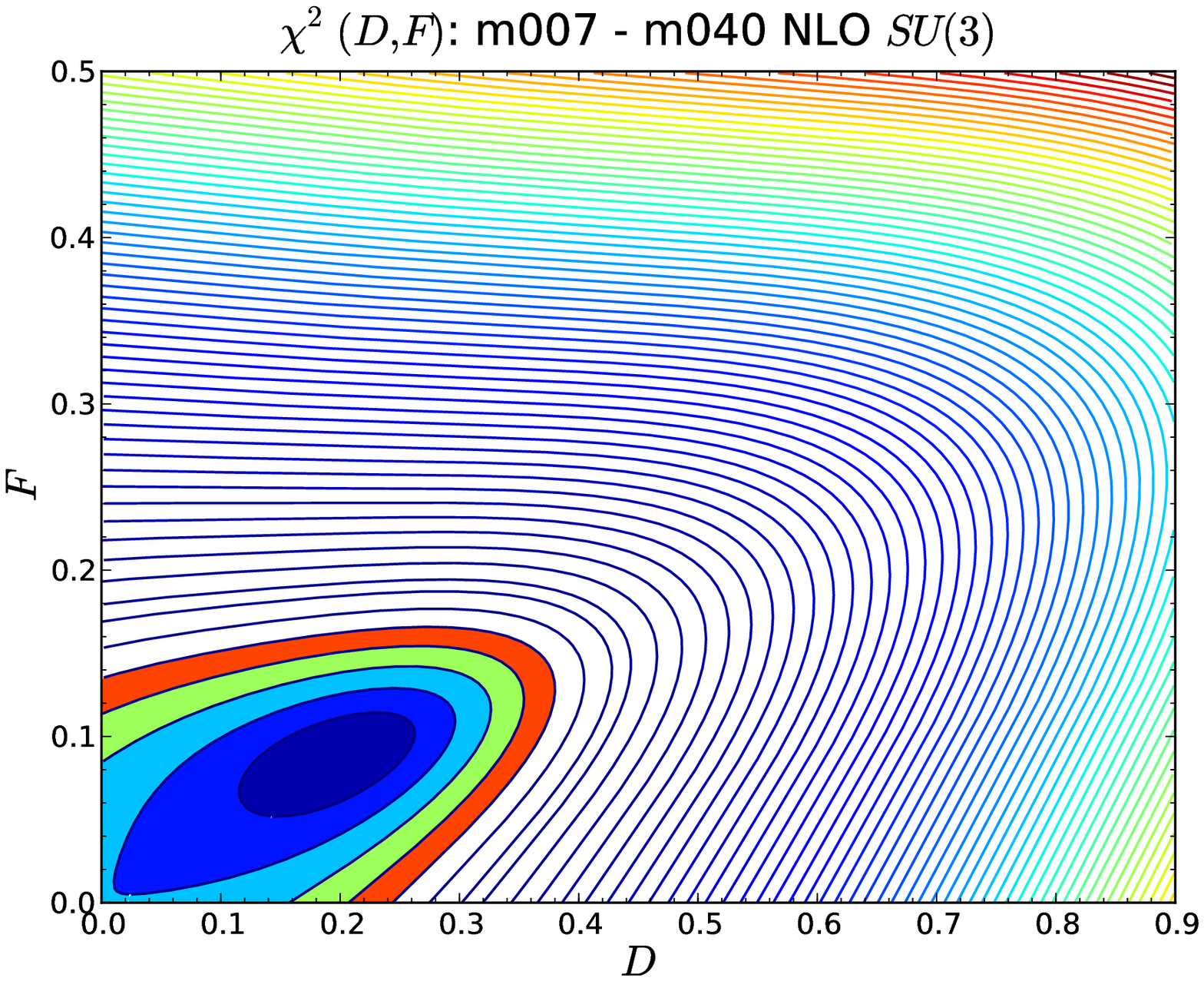}
\caption{Contour plot of $\chi^2(D,F)$ constructed using a continuum $SU(3)$ HB$\chi$PT $\chi^2$ fit to the \texttt{m007}--\texttt{m040} data sets.  The dark (blue) inner region represents $\chi^2 \lesssim \chi^2_{min} + 2.30$, the $\pm1\sigma$ confidence region for two fit parameters, $D$ and $F$.  Each successive $n^{\rm th}$ contour represents the $\pm n \sigma$ confidence region. \label{fig:chisq007040}}
\end{figure}

\begin{figure}
\includegraphics[width=8cm]{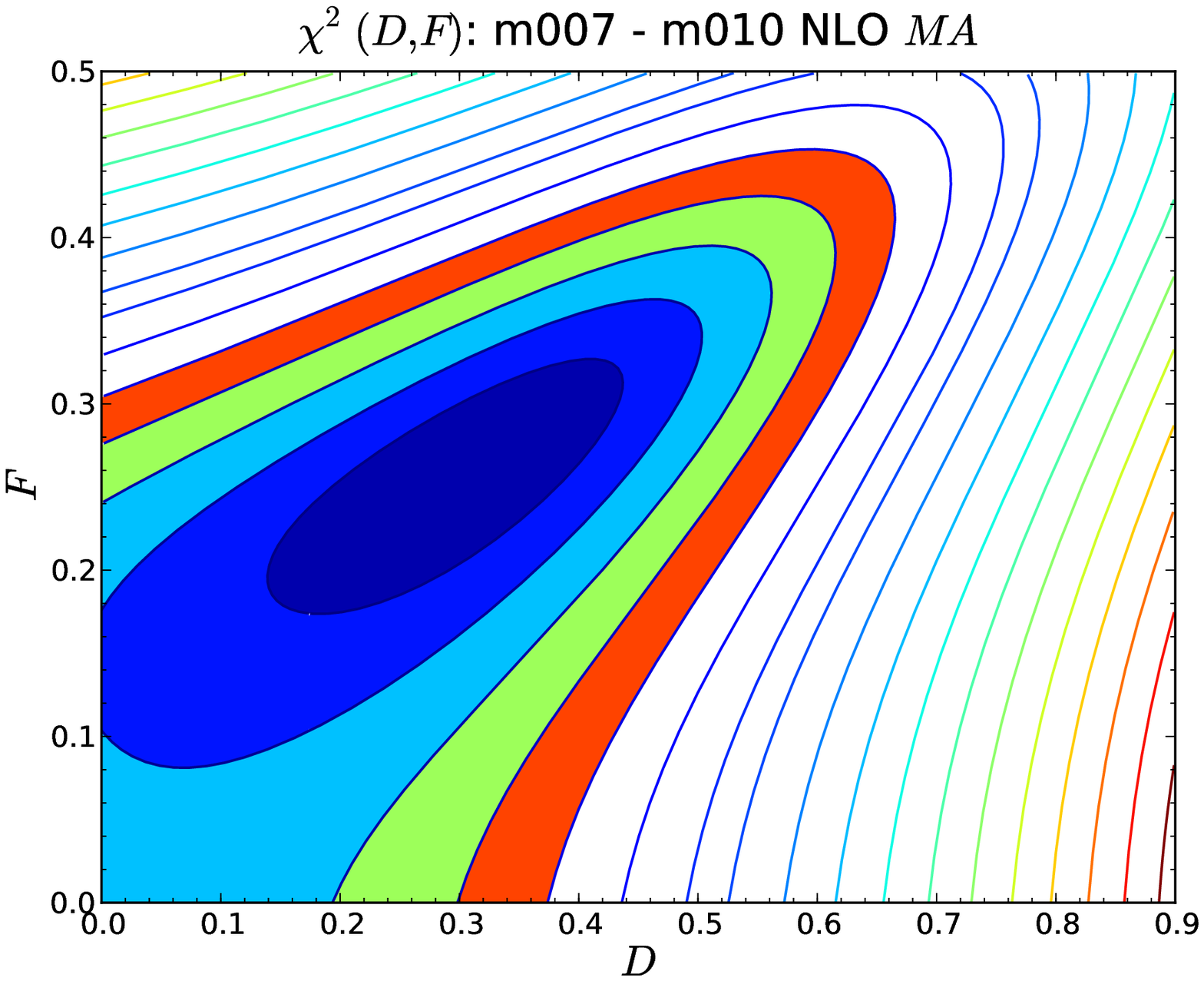}
\caption{Contour plot of $\chi^2(D,F)$ constructed using a mixed-action $SU(6|3)$ HB$\chi$PT $\chi^2$ fit to the \texttt{m007}--\texttt{m010} data sets.  The dark (blue) inner region represents $\chi^2 \lesssim \chi^2_{min} + 2.30$, the $\pm1\sigma$ confidence region for two fit parameters, $D$ and $F$.  Each successive $n^{\rm th}$ contour represents the $\pm n \sigma$ confidence region. \label{fig:chisqMA007010}}
\end{figure}

\begin{figure}
\includegraphics[width=8cm]{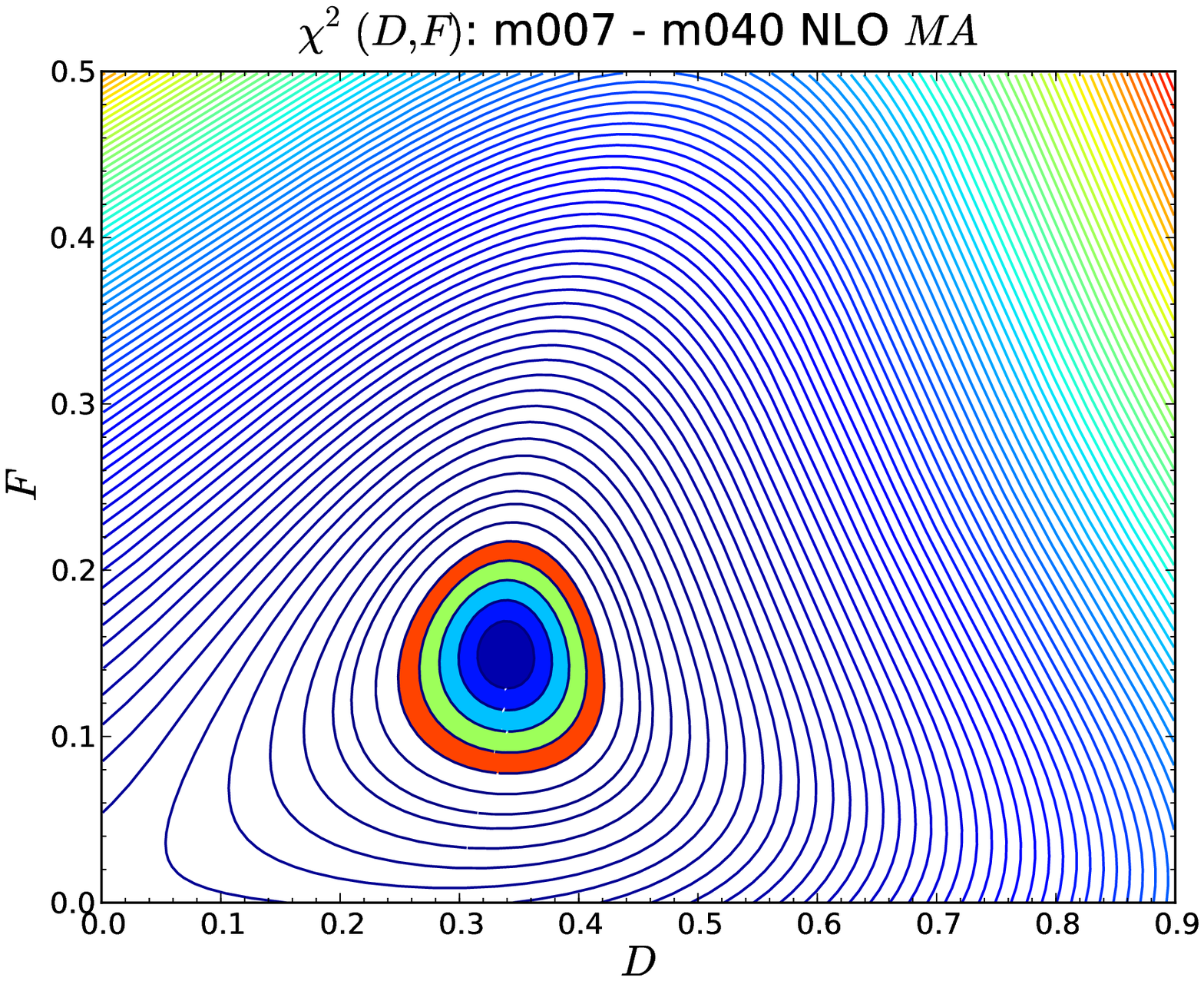}
\caption{Contour plot of $\chi^2(D,F)$ constructed using a mixed-action $SU(6|3)$ HB$\chi$PT $\chi^2$ fit to the \texttt{m007}--\texttt{m040} data sets.  The dark (blue) inner region represents $\chi^2 \lesssim \chi^2_{min} + 2.30$, the $\pm1\sigma$ confidence region for two fit parameters, $D$ and $F$.  Each successive $n^{\rm th}$ contour represents the $\pm n \sigma$ confidence region.  \label{fig:chisqMA007040}}
\end{figure}

Consistent with Refs.~\cite{WalkerLoud:2008bp,Ishikawa:2009vc}, we find the values of $D$ and $F$ are significantly smaller than those determined either phenomenologically~\cite{FloresMendieta:1998ii} or with the recent lattice determination~\cite{Lin:2007ap} (these determinations provide $D\simeq 0.72$ and $F\simeq 0.45$).  With the use of the VarPro method, we have reduced the numerical minimization to a two-dimensional problem, thus allowing us to plot the resulting $\chi^2$ as a function of $D$ and $F$.  In Fig.~\ref{fig:chisq007010}, we provide a contour plot of the resulting $\chi^2$ for the continuum fit using the lightest two ensembles.  In Fig.~\ref{fig:chisq007040}, we plot the resulting $\chi^2$ for the continuum extrapolation using the lightest five ensembles.  Figures~\ref{fig:chisqMA007010} and \ref{fig:chisqMA007040} display the same plots constructed with the mixed-action extrapolation formulae.  In all these plots, the dark (blue) central area satisfies $\chi^2 \lesssim \chi^2_{\rm min} +2.30$, the $\pm \sigma$ confidence region for two parameters.  Each successive $n^{\rm th}$ contour represents the $\pm n\sigma$ confidence region.

One observes that the mixed-action analysis returns values of $D$ and $F$ which are slightly larger than the continuum $SU(3)$ analysis, but the values are still smaller than expected.  In Table~\ref{table:R1R7Predictions}, we use the values of the LECs from the continuum analysis to compare with the experimental values of the individual baryon masses and the $1/N_c$ mass combinations in Table~\ref{table:relns} (within a few sigma, the mixed-action extrapolations are consistent).  Despite the apparent discrepancy in the values of $D$ and $F$, one observes the extrapolated values of both the baryon masses and the mass combinations, measured as a percent deviation, are in reasonable agreement with experiment.

We caution that one cannot draw strong conclusions about $SU(3)$ HB$\chi$PT from this analysis, in particular the small values of $D$ and $F$.  First, the values of $M_{K,\pi}$ used in this work are still larger than desirable for performing chiral extrapolations (in fact the strange quark mass is known to be $\sim25$\% too heavy~\cite{Aubin:2004ck}).  Second, at NLO, the baryon masses are predicted to have large $M_{K,\pi}^3$ corrections; however, this large nonanalytic contribution is not observed in the numerical results for the baryon masses themselves, see Fig.~\ref{fig:mbt}.  Consequently, there must be strong cancellations between the different orders in the chiral expansion to produce the observed results.  In order for the chiral extrapolation analysis to be consistent with both the baryon masses and the expected values of $D$ and $F$, one may need to use lighter quark masses and to include $\mc{O}(M_{K,\pi}^4)$ contributions to the masses.  Further, it is likely that a combined analysis of both the masses and the axial couplings will be required.
\begin{table}
\caption{\label{table:R1R7Predictions} Predicted baryon masses and mass relations from fit in Table~\ref{table:NLOR1R7}.  The predicted values are listed according to the heaviest light-quark mass used.  The \texttt{m030} fit is consistent with that of \texttt{m020} and \texttt{m040}.}
\begin{ruledtabular}
\begin{tabular}{|c|l|lll|}
$\mreln_i$& Exp.& \texttt{m010}& \texttt{m020}& \texttt{m040} \\
&  [\ufont{GeV}]&  [\ufont{GeV}]&  [\ufont{GeV}]&  [\ufont{GeV}] \\
\hline
$M_N$& 0.939& 1.039(31)& 1.023(12)& 1.020(8) \\
$M_\L$& 1.116& 1.159(22)& 1.145(9)& 1.142(6) \\
$M_\S$& 1.193& 1.219(22)& 1.231(9)& 1.221(7) \\
$M_\Xi$& 1.318& 1.306(16)& 1.309(7)& 1.303(5) \\
$M_\D$& 1.232& 1.376(30)& 1.454(12)& 1.427(9) \\
$M_{\S^*}$& 1.385& 1.461(28)& 1.531(10)& 1.516(8) \\
$M_{\Xi^*}$& 1.533& 1.543(27)& 1.600(10)& 1.598(7) \\
$M_\O$&1.672& 1.622(27)& 1.663(9)& 1.672(6) \\
\hline
$\mreln_1$& 175& 179(4)& 176(2)& 176(1)\\
$\mreln_2$& -9.2& -11(1)& -13.6(4)& -13.2(3)\\
$\mreln_3$& -9.03& -6.6(6)& -7.5(2)& -7.2(1)\\
$\mreln_4$& -0.21& -0.15(1)& -0.22(1)& -0.20(1)\\
$\mreln_5$& -0.21& 0.05(12)& 0.07(5)& -0.00(4)\\
$\mreln_6$& -0.73& -0.12(14)& 0.17(7)& 0.22(4)\\
$\mreln_7$& -0.092& 0.01(3)& 0.09(1)& 0.109(9)\\
\hline
$\mreln_A$& -0.024& 0.00(1)& 0.009(5)& 0.001(4)\\
$\mreln_B$& 0.0097& -0.003(8)& -0.001(4)& 0.004(3)\\
$\mreln_C$& 0.0031& -0.000(1)& -0.0033(5)& -0.0039(3)\\
\end{tabular}
\end{ruledtabular}
\end{table}

Performing the complete NNLO analysis of the octet and decuplet baryon masses introduces 19 new unknown LECs (for 30 total).  Twelve of these LECs correspond to $\mc{O}(m_q^2)$ operators of the form $\Tr ( \bar{B} \mc{M}_+ \mc{M}_+ B)$ and give rise to $M_{\pi,K,\eta}^4/\L_\chi^3$ corrections to the baryon masses.  The other seven LECs correspond to interactions of the baryons with the axial current, of the form $\Tr ( \bar{B} \mc{A} \cdot \mc{A} B)$, and give rise to baryon mass corrections of the form $M_{\pi,K,\eta}^4 \ln (M_{\pi,K,\eta}^2) /\L_\chi^3$.  The lattice data used in this work, Ref.~\cite{WalkerLoud:2008bp}, are not sufficient to precisely determine all these LECs.  Further, the number of LECs for which the mass relations are nonlinearly dependent increases to four.  In performing the full analysis, we use the VarPro method to reduce the number of LECs minimized numerically to 9.  To perform the partial NNLO analysis, we add only the $\mc{O}(m_q^2)$ operators and are able to reduce the LECs which require numerical minimization to $D$ and $F$, as in the NLO analysis.  

\begin{figure}[th]
\includegraphics[width=8cm]{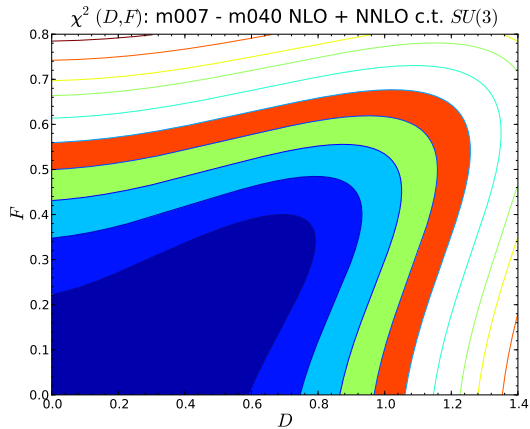}
\caption{\label{fig:chisqnnloct}Contour plot of $\chi^2(D,F)$ constructed using a continuum $SU(3)$ HB$\chi$PT $\chi^2$ fit including partial NNLO counterterms, as discussed in the text, to the \texttt{m007}--\texttt{m040} data sets.  The dark (blue) inner region represents $\chi^2 \lesssim \chi^2_{min} + 2.30$, the $\pm1\sigma$ confidence region for two fit parameters, $D$ and $F$.  Each successive $n^{\rm th}$ contour represents the $\pm n \sigma$ confidence region. }
\end{figure}

\begin{figure}[th]
\includegraphics[width=8cm]{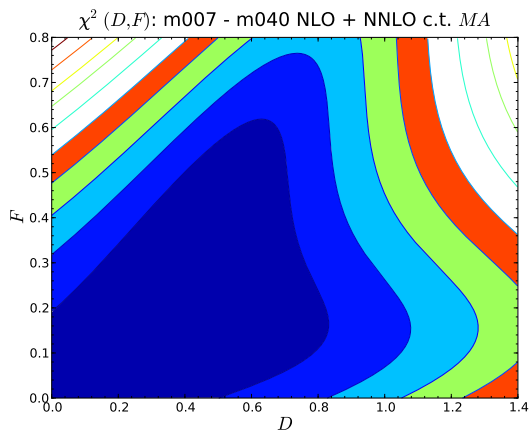}
\caption{\label{fig:chisqnnloctMA}Contour plot of $\chi^2(D,F)$ constructed using a mixed-action $SU(6|3)$ HB$\chi$PT $\chi^2$ fit including partial NNLO counterterms, as discussed in the text, to the \texttt{m007}--\texttt{m040} data sets.  The dark (blue) inner region represents $\chi^2 \lesssim \chi^2_{min} + 2.30$, the $\pm1\sigma$ confidence region for two fit parameters, $D$ and $F$.  Each successive $n^{\rm th}$ contour represents the $\pm n \sigma$ confidence region. }
\end{figure}

In Figs.~\ref{fig:chisqnnloct} and ~\ref{fig:chisqnnloctMA}, we display $\chi^2$ contours as a function of $D$ and $F$ analyzed for the continuum $SU(3)$ and the mixed-action extrapolation formulae, respectively.  In both cases, the five lightest ensembles, \texttt{m007}--\texttt{m040}, are used.
As can be seen, although the values of $D$ and $F$ are still consistent with zero, they are now also consistent with the phenomenological and lattice values, albeit with large uncertainties.  This large allowed variation of $D$ and $F$ translates into about an order of magnitude increase in the uncertainties of the predicted baryon mass relations as compared to those in Table~\ref{table:R1R7Predictions}.  The full NNLO analysis yields results in qualitative agreement with the partial NNLO minimization; however, there are more than two LECs upon which the $\chi^2$ depends nonlinearly, and thus we cannot display the corresponding contour plots.

This analysis makes it seem plausible that a combined NNLO analysis of the baryon masses and the axial couplings could provide much more stringent constraints on the values of $D$ and $F$, as well as the other LECs.  Such an analysis would allow for the first rigorous exploration of octet and decuplet properties using $SU(3)$ heavy baryon chiral perturbation theory.

\section{Conclusions}
In this work, we have explored $1/N_c$ baryon mass relations using lattice QCD.  Some of these relations were derived assuming approximate $SU(3)$ flavor-symmetry, $\mreln_1$--$\mreln_8$, while other relations were derived assuming  only $SU(2)$ flavor-symmetry, $\mreln_A$--$\mreln_D$.  In all cases, we have found that the baryon mass relations are of the size expected by both the $1/N_c$ power counting as well as the power counting in the $SU(3)$ breaking parameter, $\epsilon = (M_K^2 - M_\pi^2) / \L_\chi^2$.  For the mass relations $\mreln_A$--$\mreln_D$, it would be interesting to consider strange quark values even heavier than those used in this work.  Two of these relations become equivalent heavy-quark baryon mass relations as $m_s \rightarrow \infty$, and this exploration would probe the transition region between $m_s / \L_{\text{QCD}}  < 1 \longrightarrow m_s / \L_{\text{QCD}} > 1$.  In the charm and bottom baryon spectra, these $1/N_c$ relations have led to stringent predictions for various mass combinations.  It also would be interesting to study $\mreln_1$--$\mreln_8$, which were derived assuming approximate $SU(3)$ flavor-symmetry, as the strange quark mass is increased, to determine the value of $m_s$ at which the perturbative $SU(3)$ expansion breaks down.

Because of the definite spin and flavor transformation properties of the baryon mass relations, the discretization corrections to the mass relations also obey the $1/N_c$ and $\epsilon$ power counting, and are suppressed.  In fact, only $\mreln_1$ is subject to the leading $\mc{O}(a^2)$ mass corrections.  The suppression of these discretization errors leads to more stringent constraints on the $SU(3)$ heavy baryon chiral extrapolation of the mass relations to the physical point, and a better method for chiral extrapolation.  It allows one to rigorously test the predictions and convergence of $SU(3)$ HB$\chi$PT.

In this work, we also have performed both $SU(3)$ and mixed-action chiral extrapolations of $1/N_c$ mass relations using a variable projected $\chi^2$ minimization.  Consistent with other works, including Ref.~\cite{WalkerLoud:2008bp} (from which our numerical data are derived), we find that a NLO analysis, which includes the leading nonanalytic mass corrections, returns values of the axial couplings which are significantly smaller than expected from either phenomenology or lattice QCD.  However, the predicted values of the mass relations, as well as the octet and decuplet masses themselves, are in good agreement with experiment.  Further, we have shown that the partial inclusion of NNLO mass corrections returns values of the axial couplings $D$ and $F$ which are consistent with expectations, albeit with large error bars.  

Unfortunately, the data set we have used in this work is insufficient to precisely constrain all the LECs in the full NNLO analysis.  However, we have demonstrated that a simultaneous extrapolation of both the hyperon axial couplings as well as the $1/N_c$ mass relations would allow for a rigorous exploration of the predictions and convergence of $SU(3)$ heavy baryon $\chi$PT.  It is possible, perhaps even likely, that the strange quark is too heavy for the $SU(3)$ theory to be convergent.  This outcome, however, fails to explain the successes of flavor $SU(3)$ symmetry observed in nature in the baryon sector.  Thus,  a detailed study of this phenomena is warranted.

\begin{acknowledgements}
We would like to thank the members of the LHP Collaboration for providing their numerical baryon mass data of Ref.~\cite{WalkerLoud:2008bp}.  AWL would like to thank Kostas Orginos for helpful discussions.  The work of EEJ and AVM was supported in part by the Department of Energy under Grant No. DE-FG03-97ER40546.  The work of JWN was supported in part by the U. S Department of Energy under Grant No. DE-FG02-94ER40818.  The work of AWL was supported in by the U.S. DOE OJI Grant No. DE-FG02-07ER41527.
\end{acknowledgements}


\begin{thebibliography}{99}

\bibitem{thooft}
  G.~'t Hooft,
  Nucl.\ Phys.\  B {\bf 72}, 461 (1974).

\bibitem{witten}
  E.~Witten,
  Nucl.\ Phys.\  B {\bf 160}, 57 (1979).

\bibitem{coleman}
S.R.~Coleman, $1/N$, Presented at 1979 Int. School of Subnuclear Physics, Pointlike Structures Inside and Outside Hadrons, Erice, Italy, 1979. 


\bibitem{dm}
  R.~F.~Dashen and A.~V.~Manohar,
  Phys.\ Lett.\  B {\bf 315}, 425 (1993);
 B {\bf 315}, 438 (1993).

\bibitem{j}
  E.~E.~Jenkins,
  Phys.\ Lett.\  B {\bf 315}, 441 (1993);
 B {\bf 315}, 431 (1993);
 B {\bf 315}, 447 (1993).


\bibitem{djm1}
  R.~F.~Dashen, E.~E.~Jenkins and A.~V.~Manohar,
  Phys.\ Rev.\  D {\bf 49} (1994) 4713.
  

\bibitem{luty}
  M.~A.~Luty and J.~March-Russell,
  Nucl.\ Phys.\  B {\bf 426} (1994) 71
  
\bibitem{carone}
  C.~Carone, H.~Georgi and S.~Osofsky,
  Phys.\ Lett.\  B {\bf 322}, 227 (1994).

\bibitem{djm2}
  R.~F.~Dashen, E.~E.~Jenkins and A.~V.~Manohar,
  Phys.\ Rev.\  D {\bf 51} (1995) 3697.

\bibitem{pirjol}
  D.~Pirjol and T.~M.~Yan,
  Phys.\ Rev.\  D {\bf 57}, 1449 (1998).

\bibitem{schat}
  C.~L.~Schat, J.~L.~Goity and N.~N.~Scoccola,
  Phys.\ Rev.\ Lett.\  {\bf 88}, 102002 (2002).

\bibitem{three}
  D.~Pirjol and C.~Schat,
  Phys.\ Rev.\  D {\bf 67}, 096009 (2003).


\bibitem{ejreview}
  E.~E.~Jenkins,
  Ann.\ Rev.\ Nucl.\ Part.\ Sci.\  {\bf 48} (1998) 81.

\bibitem{amreview}
  A.~V.~Manohar, {\sl Large N QCD,} Les Houches Summer School in Theoretical Physics, Session 68: Probing the Standard Model of Particle Interactions, Les Houches, France, 1997,
  arXiv:hep-ph/9802419.
  
\bibitem{jl}
  E.~E.~Jenkins and R.~F.~Lebed,
  Phys.\ Rev.\  D {\bf 52} (1995) 282.
  
\bibitem{h}
  E.~E.~Jenkins,
  Phys.\ Rev.\  D {\bf 54} (1996) 4515,
D {\bf 55} (1997) R10,
  D {\bf 77} (2008) 034012.


\bibitem{Durr:2009ma}
  S.~Durr {\it et al.},
  Science {\bf 322}, 1224 (2008).

\bibitem{flores}
  R.~Flores-Mendieta et al.,
  Phys.\ Rev.\  D {\bf 62}, 034001 (2000).

\bibitem{Teper:1998te}
  M.~J.~Teper,
  Phys.\ Rev.\  D {\bf 59}, 014512 (1998).
\bibitem{teper}
  M.~J.~Teper, {\sl Large N,}
  arXiv:0812.0085 [hep-lat].

\bibitem{Kennedy:2004ae}
  A.~D.~Kennedy,
  Nucl.\ Phys.\ Proc.\ Suppl.\  {\bf 140}, 190 (2005).

\bibitem{Aoki:2008sm}
  S.~Aoki {\it et al.}  [PACS-CS Collaboration],
  arXiv:0807.1661 [hep-lat].

\bibitem{wise}
  E.~E.~Jenkins, A.~V.~Manohar and M.~B.~Wise,
  Phys.\ Rev.\ Lett.\  {\bf 75} (1995) 2272.
  
\bibitem{Beane:2006pt}
  S.~R.~Beane, K.~Orginos and M.~J.~Savage,
  Phys.\ Lett.\  B {\bf 654}, 20 (2007)
  [arXiv:hep-lat/0604013].

\bibitem{Jenkins:1990jv}
  E.~E.~Jenkins and A.~V.~Manohar,
  Phys.\ Lett.\  B {\bf 255}, 558 (1991);
  {\bf 259}, 353 (1991).

\bibitem{georgi}
  A.~Manohar and H.~Georgi,
  Nucl.\ Phys.\  B {\bf 234} (1984) 189.
  
\bibitem{WalkerLoud:2008bp}
  A.~Walker-Loud {\it et al.},
  Phys.\ Rev.\  D {\bf 79}, 054502 (2009).

\bibitem{Kaplan:1992bt}
  D.~B.~Kaplan,
  Phys.\ Lett.\  B {\bf 288}, 342 (1992).

\bibitem{Shamir:1993zy}
  Y.~Shamir,
  Nucl.\ Phys.\  B {\bf 406}, 90 (1993).

\bibitem{Furman:1994ky}
  V.~Furman and Y.~Shamir,
  Nucl.\ Phys.\  B {\bf 439}, 54 (1995).

\bibitem{Orginos:1998ue}
  K.~Orginos and D.~Toussaint  [MILC collaboration],
  Phys.\ Rev.\  D {\bf 59}, 014501 (1998).

\bibitem{Orginos:1999cr}
  K.~Orginos, D.~Toussaint and R.~L.~Sugar  [MILC Collaboration],
  Phys.\ Rev.\  D {\bf 60}, 054503 (1999).

\bibitem{Bernard:2001av}
  C.~W.~Bernard {\it et al.},
  Phys.\ Rev.\  D {\bf 64}, 054506 (2001).

\bibitem{Bazavov:2009bb}
  A.~Bazavov {\it et al.},
  arXiv:0903.3598 [hep-lat].

\bibitem{Beane:2008dv}
  S.~R.~Beane, K.~Orginos and M.~J.~Savage,
  Int.\ J.\ Mod.\ Phys.\  E {\bf 17}, 1157 (2008).

\bibitem{Aubin:2004ck}
  C.~Aubin {\it et al.}  [HPQCD Collaboration and MILC Collaboration and
                  UKQCD Collaboration],
  Phys.\ Rev.\  D {\bf 70}, 031504 (2004).

\bibitem{Aubin:2004wf}
  C.~Aubin {\it et al.},
  Phys.\ Rev.\  D {\bf 70}, 094505 (2004).

\bibitem{pdg}
C.~Amsler {\it et al.}  [Particle Data Group],
  Phys.\ Lett.\  B {\bf 667} (2008) 1.

\bibitem{Beane:2009ky}
  S.~R.~Beane {\it et al.},
  Phys.\ Rev.\  D {\bf 79}, 114502 (2009)
  [arXiv:0903.2990 [hep-lat]].

\bibitem{WalkerLoud:2008pj}
  A.~Walker-Loud,
  arXiv:0810.0663 [hep-lat].
  
\bibitem{Ishikawa:2009vc}
  K.~I.~Ishikawa {\it et al.}  [PACS-CS Collaboration],
  arXiv:0905.0962 [hep-lat].

\bibitem{FloresMendieta:1998ii}
  R.~Flores-Mendieta, E.~E.~Jenkins and A.~V.~Manohar,
  Phys.\ Rev.\  D {\bf 58}, 094028 (1998);
   J.~Dai, {\sl et al.},
  Phys.\ Rev.\  D {\bf 53}, 273 (1996).


\bibitem{Lin:2007ap}
  H.~W.~Lin and K.~Orginos,
  Phys.\ Rev.\  D {\bf 79}, 034507 (2009).

\bibitem{Jenkins:1991ts}
  E.~E.~Jenkins,
  Nucl.\ Phys.\  B {\bf 368}, 190 (1992).

\bibitem{Lebed:1994gt}
  R.~F.~Lebed and M.~A.~Luty,
  Phys.\ Lett.\  B {\bf 329}, 479 (1994).

\bibitem{Borasoy:1996bx}
  B.~Borasoy and U.~G.~Meissner,
  Annals Phys.\  {\bf 254}, 192 (1997).

\bibitem{WalkerLoud:2004hf}
  A.~Walker-Loud,
  Nucl.\ Phys.\  A {\bf 747}, 476 (2005).

\bibitem{Lebed:1993yu}
  R.~F.~Lebed,
  Nucl.\ Phys.\  B {\bf 430}, 295 (1994).

\bibitem{Tiburzi:2004rh}
  B.~C.~Tiburzi and A.~Walker-Loud,
  Nucl.\ Phys.\  A {\bf 748}, 513 (2005).

\bibitem{Chen:2001yi}
  J.~W.~Chen and M.~J.~Savage,
  Phys.\ Rev.\  D {\bf 65}, 094001 (2002).

\bibitem{Tiburzi:2005is}
  B.~C.~Tiburzi,
  Phys.\ Rev.\  D {\bf 72}, 094501 (2005).

\bibitem{Chen:2007ug}
  J.~W.~Chen, D.~O'Connell and A.~Walker-Loud,
  JHEP {\bf 0904}, 090 (2009).

\bibitem{Orginos:2007tw}
  K.~Orginos and A.~Walker-Loud,
  Phys.\ Rev.\  D {\bf 77}, 094505 (2008).

\bibitem{VarPro}
  G.~Golub and V.~Pereyra,
  Inverse Problems, {\bf 19}, R1-R26 (2003).  
    
\end{thebibliography}
\end{document}